\documentclass{aa}  

\usepackage{txfonts}
\usepackage{amsmath}	
\usepackage{xcolor}
\usepackage{natbib}
\usepackage{hyperref}
\usepackage{pifont}
\usepackage{graphicx,dblfloatfix}
\usepackage{adjustbox}
\usepackage{booktabs}

\definecolor{verde}{rgb}{0.,0.56,0.}
\definecolor{lightblue}{rgb}{0.1,0.6,0.93}
\definecolor{mblue}{rgb}{0, 0.5, 0.815}
\definecolor{purple1}{rgb}{0.5,0,0.87}

\renewcommand*{\bibfont}{\small}
\hypersetup{colorlinks,linkcolor={cyan},citecolor={blue},urlcolor={purple}} 

\newcommand{\sextractor}{\texttt{SExtractor}}

\newcommand{\dg}{MATLAS-2019}
\newcommand{\matlasEffRad}{$17.\!^{\prime\prime}2$}
\newcommand{\finalPopulation}{$33\pm3$}
\newcommand{\totalMass}{$(1.14\pm0.1)\times10^{11} M_\odot$}
\newcommand{\gaussianDist}{$19.9\pm 0.8$}
\newcommand{\kingDist}{$20.1\pm 0.8$}
\newcommand{\meanDist}{$20.0\pm 0.9$}

\newcommand{\stellarMass}{$8.6^{+1.5}_{-1.3} \times 10^7$ M$_\odot$}
\usepackage{dblfloatfix}
\usepackage{float}

\begin{document} 
   \title{The mysterious Globular Cluster population of \dg}
   \author{Sergio Guerra Arencibia \inst{1,2}
           \and
           Mireia Montes \inst{3,1,2}
           \and
           Giulia Golini \inst{1, 2}
           \and
           Ignacio Trujillo \inst{1,2}
          }

\institute{Instituto de Astrof\'isica de Canarias, c/ V\'ia L\'actea s/n, 38205 La Laguna, Tenerife, Spain 
\and
Departamento de Astrof\'isica, Universidad de La Laguna, 38206 La Laguna, Tenerife, Spain
\and
Institute of Space Sciences (ICE, CSIC), Campus UAB, Carrer de Can Magrans, s/n, 08193 Barcelona, Spain
}
\date{ }
\keywords{Galaxies: dwarf -- dark matter -- galaxies: photometry -- galaxies: structure -- galaxies: formation}

  \abstract
   { \dg{} (also known as NGC5846-UDG1) has attracted significant attention due to the ongoing debate surrounding its Globular Cluster (GC) population, with several studies addressing the issue yet reaching little consensus. 
   In this paper we take advantage of HST's multi-wavelength coverage (\textit{F475W}, \textit{F606W} and \textit{F814W} observations) with the addition of deep $u$-band imaging from \textit{Gran Telescopio de Canarias}, to perform the most detailed study and estimation to date of the GC population of the ultra-diffuse galaxy \dg. The improved constraints provided by the combination of high spatial resolution and better coverage of the GC spectral energy distribution has allowed us to obtain a clean sample of GCs in this galaxy. We report a number of \finalPopulation{} GCs in \dg, supporting the previous lower estimates for this galaxy. The GC population of this galaxy is highly concentrated with $\sim 80\%$ of the GCs inside the effective radius ($R_e$) of the galaxy and the GC half-number radius $R_{e,GC}$ is $0.7\times R_e$. Using the GC-Halo mass relation, we estimate a halo mass for \dg{} of \totalMass. The GC luminosity function and the distribution of effective radii of the GCs favour a distance to the galaxy of \meanDist{} Mpc. In agreement with previous findings, we find that the distribution of GCs is highly asymmetric even though the distribution of stars in the galaxy is symmetric. This suggests that assumptions about the symmetry of the GC distribution may be incorrect when used to calculate the number of GCs with such low statistics. }
   \keywords{dark matter --
                galaxies: formation --
                galaxies: evolution --
                galaxies: individual (\dg)
               }

   \maketitle

\section{Introduction}

Ultra-diffuse galaxies (UDGs) are a subset of dwarf galaxies with two identifying characteristics: a faint central surface brightness ($\mu_g(0) \gtrsim 24$ mag arcsec$^{-2}$) and a large effective radius ($R_e \gtrsim 1.5$ kpc; \citealt{VanDokkum2015}). They have been found in a wide range of environments: clusters (e.g. Coma: \citealt{VanDokkum2015}, Fornax: \citealt{munoz2015}, Virgo: \citealt{Lim2020}, Perseus: \citealt{Marleau2025}), groups (e.g. \citealt{Roman_Trujillo2017, Trujillo2017, Galinsky2023, fielder2024}) and the field \citep[e.g.][]{Greco2018, Roman_Beasley_2019, Marleau2021, Jones2023}. Two main hypothesis have been proposed to explain the origin of objects with such properties. The first is that they are normal dwarf galaxies that underwent some process that `puffed' them up: high initial spins \citep{AmoriscoLoeb2016}, stellar feedback \citep{diCintio2016}, early mergers \citep{wright2021}, tidal heating of dwarf galaxies \citep{fielder2024}, effects of clusters tides \citep{sales2020} or tidal disruption \citep{zemaitis2023}.
The second hypothesis is that they were going to host a larger stellar content but were prematurely quenched and ceased to form stars. Consequently, they are now embedded in dark matter haloes that are more massive than expected given their stellar mass \citep{VanDokkum2015, vanDokkum2016, BeasleyTrujillo2016, Toloba2018}. 

One way to distinguish between these formation scenarios is to infer the mass of the dark matter halos, as it will give us valuable insights into the nature of these objects. However, using direct methods like stellar velocity dispersion or rotation curves is usually not feasible for such faint systems and indirect measurements of the masses need to be employed. For example, the number of Globular Clusters ($N_{GC}$) appears to correlate with the total mass of the galaxy \citep[e.g.][]{Blakeslee1997, SptilerAndForbes2009, Hudson2014, Harris2017, Forbes2018, Le2025}. If this relation also holds for UDGs then this will provide a relatively easy way to infer the amount of dark matter of these systems. 

Studies of UDGs have found both rich GCs systems \citep{van_Dokkum2017, Forbes2020, Muller2021, Danieli2022, Marleau2024} and systems in line with typical dwarf galaxies of the same stellar masses \citep{Amorisco2018, Saifollahi2020, Marleau2021}. But there is one particular galaxy that stands out for its abundant GC system. This object, NGC5846-UDG1 or \dg, was identified as a \textit{very low surface brightness} (VLSB) galaxy with the name NGC 5846-156 by \citet{Mahdavi2005} but later re-classified as an UDG in \cite{Forbes2019} and \cite{Poulain2021}. It has an effective radius ($R_e$) of \matlasEffRad{} \citep{Muller2020} and its systemic velocity ($2156\pm9$ km$\,$s$^{-1}$, \citealt{Muller2020}) is consistent with the velocity of the NGC 5846 group, which suggests that it is a member of the group. 

\citet{Forbes2020b} estimated a number of $\sim$45 GCs using ground-based imaging from the VEGAS survey, compatible with \citet{Danieli2022}, who found $54\pm9$ GCs, using high-resolution \textit{Hubble Space Telescope} (HST) imaging (WFC3; \textit{F475W} and \textit{F606W} bands), more than any previously known galaxy of similar properties. In contrast, \citet{Muller2021}, using another set of observations from the HST (ACS; \textit{F606W} and \textit{F814W} bands), found a population of $26\pm6$ using a Bayesian approach ($36\pm6$ using a non-Bayesian approach) which places \dg's GC system in line with the systems of other UDGs. A more recent study from \cite{Marleau2024} finds, using also HST (ACS; \textit{F606W} and \textit{F814W}), $38\pm7$ GCs for \dg, somewhere in between previous studies. 

In addition to the disagreement in the number of GCs, the distance to the galaxy is still an open question. Using the surface brightness fluctuations (SBF) technique, \cite{Danieli2022} measured a distance of $21 \pm 5$ Mpc to \dg{} but ended up using the group's distance due to the low signal to noise ratio of the measurement (i.e. $26.5$ Mpc from the weighted average distance of the members from Cosmicflows-3 catalogue \citealt{Tully2016, Kourkchi2017, Danieli2022}). On the other hand \cite{Muller2021}, using the peak of the GC Luminosity Function (GCLF), finds a distance of $20.7^{+2.3}_{-2.1}$ Mpc. 

In this paper, we take advantage of all the HST's high-resolution, multi-band imaging to revise the estimation of the number of GCs of \dg. The improved constrains provided by the high spatial resolution and better sampling of the spectral energy distribution of these objects are necessary to minimise contamination from foreground and background sources. We also use deep data from the \textit{Gran Telescopio Canarias} (GTC) to refine the identification of GCs and to characterise the stellar body of this galaxy. We compare our results with previous estimates of the GC population from the literature and explore its implications for our understanding of UDG formation. Throughout this work we adopt a standard cosmological model with the following parameters: $H_0=70$ km s$^{-1}$ Mpc$^{-1}$, $\Omega_m=0.3$ and $\Omega_\Lambda=0.7$. 
All magnitudes are given in the AB system unless otherwise specified.

\section{Data}

The data used in this paper come from two different facilities: the Hubble Space Telescope (HST) and the 10.4m \textit{Gran Telescopio de Canarias} (GTC). The details of the observations are described below.

\subsection{Hubble Space Telescope imaging}

\dg{} was observed with HST as part of two different sets of observations. One using the Advance Camera for Surveys (ACS) and the other using the Wide Field Camera 3 (WFC3). Both datasets were retrieved from the MAST archive\footnote{\url{https://mast.stsci.edu/portal/Mashup/Clients/Mast/Portal.html}}.

\subsubsection{Advanced Camera for Surveys}

The ACS images were obtained using the Wide Field Channel (WFC) as part of the program GO-16082 (PI: Müller). The galaxy was observed in the \textit{F606W} and \textit{F814W} bands, although in this paper we will only use \textit{F814W}. The reason for this is deeper \textit{F606W} data exist taken with the WFC3. The total exposure time of the images is $1030$ seconds for both bands. We used the charge-transfer efficiency (CTE) corrected images (\textit{.drc.fits} files) produced by the standard pipeline.

\subsubsection{Wide Field Camera 3}

The images from the WFC3 were obtained using the UVIS channel as part of the program GO-16284 (PI: Danieli). The galaxy was observed in the \textit{F475W} and \textit{F606W} bands, with a total exposure time of $2349$ and $2360$ seconds respectively. The CTE corrected images (\textit{.drc.fits} files) produced by the standard pipeline are used as well.

\subsection{GTC images}
In order to improve our identification of GCs and study the light distribution of \dg{}, we obtained deep optical and multi-band imaging GTC, using the upgraded camera \textit{Optical System for Imaging and low-Intermediate-Resolution Integrated Spectroscopy} (OSIRIS+). The galaxy was observed between the 18 and 26 of April 2023 in the \textit{u}, \textit{g} and \textit{r} filters as part of the program GTC57-22B (PI: Trujillo). The final exposure times on-source are 5580, 5940 and 5760 seconds, respectively. The observational strategy and data reduction of the observations are described in Appendix \ref{appendix:data_reduction}.

\subsection{Image depth}
\label{sec:DataDepth}

The aim of this work is to obtain the most accurate sample of GC candidates of \dg. To achieve this, we first need to estimate the limiting depth of our images which characterises the magnitude to which is reasonable to trust our detections.
Since we are interested in characterising the limits for detecting GCs, we compute this limiting depth on our images using apertures based on the full-width-at-half-maximum (FWHM) that the spectroscopically confirmed GCs have in the images (see Sec. \ref{sec:GCprofiles}). Thus, we randomly place, after masking all the sources in the images, 20000 apertures of diameter twice the FWHM of the GC profiles in each image. That is: $0\farcs216$ (5.4 px) for the WFC3 data, $0\farcs270$ (5.4 px) for the ACS data, and $2\farcs2$ (8.6 px) for the OSIRIS+ data. We calculate the resulting distributions and its $5\sigma$ values, which are listed in Table \ref{tab:depths}. 

\begin{table}[h]
\centering
\caption{5$\sigma$ limiting depths using apertures twice the FWHM of the GC profiles for each of the images used to identify the GC candidates of \dg{}.}
\begin{adjustbox}{width=6.5cm}
\begin{tabular}{c cccc}
\toprule
Instrument & Filter & $D_{aperture}$ & $mag_{limit}$\\
 & & (arcsec) & (mag) &    \\
\midrule
WFC3 & \textit{F475W} & 0.216 & 27.05 \\ 
WFC3 & \textit{F606W} & 0.216 & 27.10 \\
ACS  & \textit{F814W} & 0.270 & 26.50 \\
OSIRIS & \textit{u}   & 2.20 & 25.75 \\
\bottomrule
\end{tabular}
\end{adjustbox}
\label{tab:depths}
\end{table}

Due to the difference in resolution and depth between the HST and GTC data, HST imaging will be used to detect and select the GC candidates of this galaxy. The OSIRIS+ images will be used to characterize the stellar body of the galaxy, while the $u$-band will also be used to clean up the initial catalogue of GC candidates. In Table \ref{tab:brightLimits} details of the OSIRIS+ data are provided.

\begin{table}[h]
\centering
\caption{Characteristics of the OSIRIS+ ultra-deep imaging of MATLAS-2019.}
\begin{adjustbox}{width=7.5cm}
\begin{tabular}{c cccc}
\toprule
Band & $t_{exposure}$ & $\mu_{limit}$ ($3\sigma, 10^{\prime\prime}\times10^{\prime\prime})$  & PSF FWHM \\
 & & (mag/arcsec$^2$) & (arcsec)   \\
\midrule
u & 1h 33min &  30.35 & 1.1 \\ 
g & 1h 39min &  31.22 & 0.8 \\
r & 1h 36min &  30.86 & 1.0 \\
\bottomrule
\end{tabular}
\end{adjustbox}
\label{tab:brightLimits}
\end{table}

\section{The globular cluster system of \dg} 

The task of identifying GCs in galaxies is not trivial. Foreground and background contamination makes it difficult to characterize the overall GC system of the host galaxy with accuracy. Recent works have shown the increased effectiveness of identifying GCs using more information from their spectral energy distributions \citep[e.g.][]{Montes2014, Munoz2014} to help separate them from contaminants. \citet{Muller2020} already identified 11 GCs in \dg{} using spectroscopy, and \citet{Haacke2025} further extended the sample of spectroscopically confirmed GCs to 20. Here, we aim to find objects that are similar (both in morphology and colours) to the spectroscopically confirmed GCs, taking into account sources that may have been too faint for spectroscopic follow-up or that are located outside the coverage of the spectroscopic observations. Note that the faintest spectroscopically confirmed GCs in \citealt{Haacke2025} are well below $2\sigma$ from the expected GCLF, see Sec. \ref{sec:finalnumber}. 

\subsection{Characterizing the light profiles of the GCs} \label{sec:GCprofiles}
At the distance of \dg, the resolution of HST allows to resolve the GCs of this galaxy, i.e. objects with an effective radius of $\sim3$ pc. The FWHM of the Point Spread Function (PSF) of the images (measured directly from stars both in WFC3 and ACS) is $\sim 2.2$ px  while the typical FWHM of a spectroscopically confirmed GC is $\sim 2.7$ px (see below). Therefore, it is crucial to study the light profiles of the spectroscopically confirmed GCs in order to define the parameters needed to perform accurate photometry: aperture, aperture correction and distance from the centre of the GC to perform the local background estimation.

\subsubsection{GC 1-D radial profiles}\label{Sec:RadialProf}

\begin{figure}[h]
    \centering
    \includegraphics[width =0.8\columnwidth]{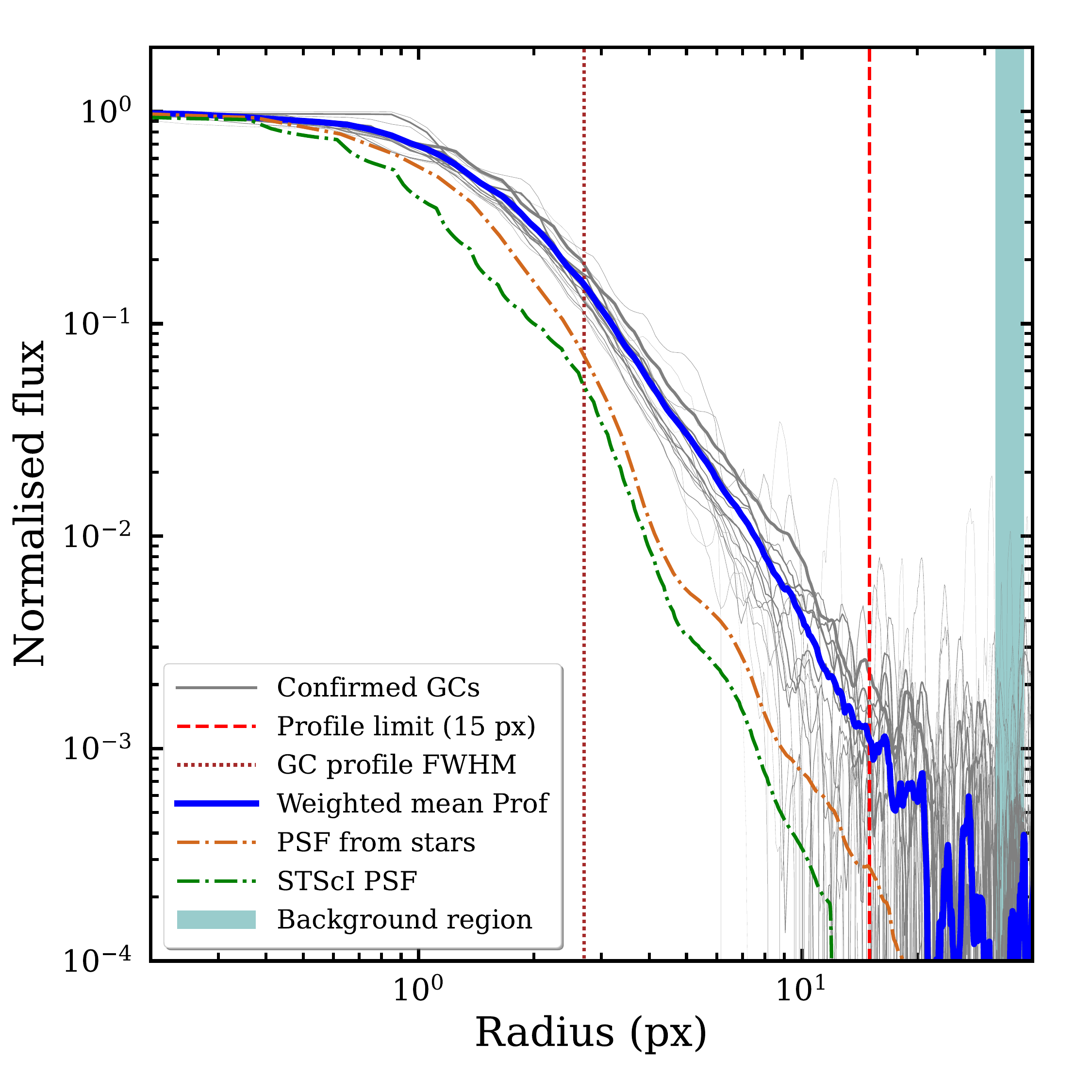}
    \caption{GC radial profiles in the WFC3/\textit{F606W} band for the 20 spectroscopically confirmed GCs (grey). The blue line shows the S/N weighted mean profile of the GCs. We also plot the profile of the empirical PSF provided by the STScI in green, and the profile of the PSF obtained directly from stars of the images in orange. The vertical dashed red line (0\farcs6, equivalent to 15 px in WFC3 imaging) indicates where the profile is truncated, the vertical dotted brown line shows the FWHM of the mean GC profile, and the light blue region indicates where the background has been estimated.}
    \label{fig:gc606Profile}
\end{figure}

Light profiles of the spectroscopically confirmed GCs are derived in sufficiently large stamps (81px $\times$ 81px), which allow us to explore the profiles until they fade away in the noise. On these stamps we measure the centroid of the objects using \texttt{centroid\_2dg} from the \texttt{photutils} package \citep{larrybradley2023}. We then characterise the local background in order to subtract it from the stamps. To do this, we explore different radii and widths for the background annulus, looking to minimise the root mean square (RMS) of the background region for each GC profile. The minimum RMS on the background region is found using an annulus with a width of 6 px ($0\farcs24$ in WFC3 and $0\farcs3$ in ACS imaging) and with an inner radius of diameter 64 px ($2\farcs56$) for the WFC3, and 60 px ($3\farcs00$) for the ACS. Once we have the background subtracted stamps, we derive the radial profiles using the \texttt{RadialProfile} routine from \texttt{photutils}. We compute the average GC profile, weighting the profiles at each radius by their total fluxes (i.e. $w_i = F_i/F_{max}$). Fig.~\ref{fig:gc606Profile} shows the radial profiles for the 20 confirmed GCs from \citet{Muller2020} and \citet{Haacke2025} (grey), the S/N-weighted mean profile (blue), the region where the background is estimated (light blue region), the profile of the empirical PSF provided by STScI\footnote{e.g. for WFC3: \url{https://www.stsci.edu/hst/instrumentation/wfc3/data-analysis/psf}} (green), and the profile of the PSF obtained directly from stars of the images (orange). All profiles are normalised to the unit at the centre. Finally, to define the limiting radius of the profiles, we truncate the profiles when they reach 5 times the value of the standard deviation of the background. To avoid any colour bias, we use the same limit for all the bands, which is the shallowest i.e. 0\farcs60 arcsec in \textit{F814W}.

\subsubsection{PSF of the images} \label{sec:psfImages}

The PSF has been modelled in a similar way as the mean profile of the GCs (see previous section), but this time combining bright, non-saturated, and non-contaminated stars from the images. Thus, we derive the 1-D radial profiles of the stars, and obtain the mean profile weighting them by their total fluxes at each radius.

In order to be more robust in our analysis of the PSF, we will take both our PSF model and the empirical PSFs provided by STScI into account (more details in Sec. \ref{sec:AperAndCorrect}). Thus, the PSF built with stars from the image has a FWHM of 2.2 px and the PSF given by STScI has a FWHM of 1.75 px.

\subsubsection{Photometric and aperture corrections} \label{sec:AperAndCorrect}

After obtaining the weighted mean radial profile of the GCs in each of the filters, we derive the photometric apertures. We defined the apertures here to have a diameter of 2 times the FWHM of the GCs. We estimate this FWHM from the mean GC profile. The FWHM is 2.7 px, both in the WFC3 and ACS images.

We also use the GC radial profiles to derive the aperture corrections up to infinity, in order to obtain the most accurate photometry of the GCs. We divide this step into two parts ---defined in terms of diameters--- from 2$\times$FWHM to 1\farcs2 and from 1\farcs2 to infinity.
The first correction, from 2$\times$FWHM to 1\farcs2, is derived from the mean GC profile as follows. We compare the amount of flux enclosed in 2$\times$FWHM with the amount of total flux of the profile up to 1\farcs2, which gives us the first correction. 

For the second step, the correction to infinity, we need to extend the GC radial profile derived above, i.e. we need to estimate the slope in the outer parts of the GC profile. The slope of the outer parts of the GC profiles at large radii is assumed to be the same to that of the PSF in the outermost region. Therefore, we fit a power law to the outer parts (from 3 px to 10 px) of the empirical PSFs provided by STScI and our PSF constructed from stars in the images. The slope we use is the average of the two derived slopes, while the difference between them is included as the uncertainty of the correction. Once we have radially extended our GC profiles, we calculate the corrections to infinity. 
Table \ref{tab:corrections} gives the corrections for the two steps, which are applied as $m_{corr} = m_{ini} + m_{corr1} + m_{corr2}$. 

\begin{table}[h]
\centering
\caption{Photometric corrections based on the properties of the GCs in the images.}
\begin{adjustbox}{width=8cm}
\begin{tabular}{c cccc}
\toprule
Instrument & Filter & Profile corr. (1st) & Corr. to infinity (2nd)\\
 & & (0 to 0\farcs60) & (0\farcs60 to infinity) &    \\
& & (mag) & (mag) &    \\

\midrule
WFC3 & \textit{F475W} & -0.62 & -0.08 $\pm$ 0.02 \\ 
WFC3 & \textit{F606W} & -0.62 & -0.06 $\pm$ 0.01 \\
ACS  & \textit{F814W} & -0.57 & -0.10 $\pm$ 0.01 \\
\bottomrule
\end{tabular}
\end{adjustbox}
\label{tab:corrections}
\end{table}

\subsection{GC pre-selection} \label{sec:GCSelection}

\begin{figure*}
    \centering
\includegraphics[width=16cm,trim=5 5 5 20,clip]{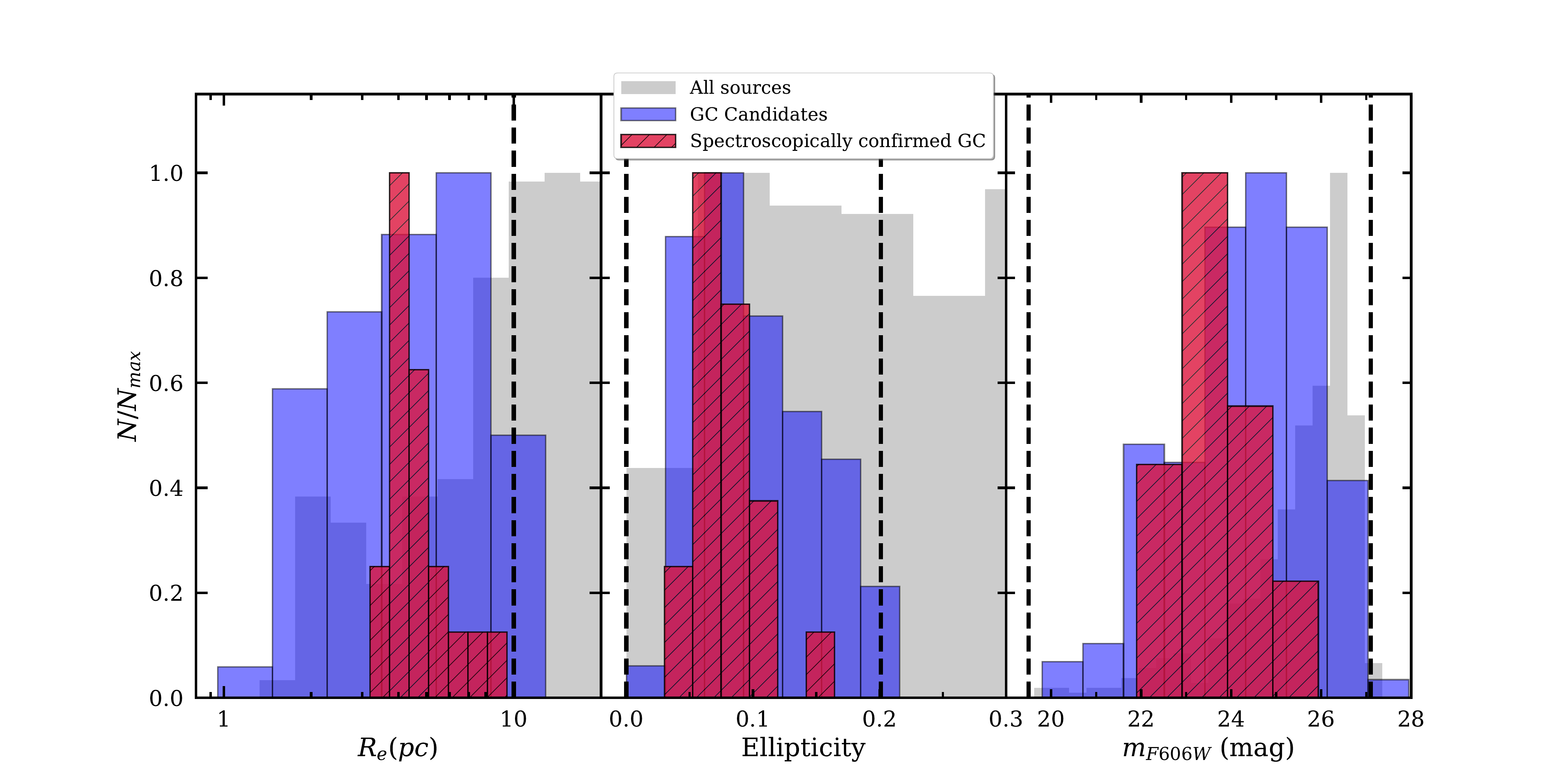}
    \caption{Normalised histograms of the effective radius (in pc, assuming a distance of 20.7 Mpc), ellipticity and $F606W$ magnitude of different samples of sources; $N$ is the number of sources in each bin and $N_{max}$ the maximum of their respective histograms. The grey histogram shows all the sources of the catalogues in grey (446 sources), the red hatched histogram the 20 spectroscopically confirmed GCs and the blue histogram the sources which fulfil the selection criteria (128 sources). The vertical dashed lines indicate the ranges of the GC selection, in each of the parameters.}
    \label{fig:ReEllipMagHists}
\end{figure*}

Once we have defined the parameters for performing photometry, we can build a preliminary catalogue of candidate GCs. As a first step, we take advantage of the high-resolution \emph{HST} data to select them based on their morphology. First, we build catalogues of objects by running \texttt{SExtractor} \citep{BertinArnouts1996} in the HST images\footnote{As a check, we repeated the analysis, this time subtracting the galaxy’s diffuse light using unsharp masking. The results did not change. This is because the galaxy is very diffuse and almost transparent so it does not hide any GC.}. To perform the photometry, we use the python package \texttt{photutils} with apertures of diameter $2\times$FWHM of the confirmed GCs, estimating the background and performing the corrections described in Sec. \ref{Sec:RadialProf} and Sec. \ref{sec:AperAndCorrect}. The Galactic extinction correction is applied afterwards ($A_{F475W}$ = 0.171, $A_{F606W}$ = 0.131, $A_{F814W}$ = 0.081)\footnote{\url{https://ned.ipac.caltech.edu/extinction\_calculator} \label{nedExtinctionCalculator}}. We match the catalogues in the three HST filters (\textit{F475W}, \textit{F606W} and \textit{F814W}) using their sky position with a tolerance based on the typical extension of the brightest GCs of the galaxy (i.e. $0.5$ arcsec).

We identify GC candidates based on their effective radius ($R_e$), ellipticity, and magnitude in the \textit{F606W} band ($m_{F606W}$). This pre-selection criteria is modelled upon the properties of the confirmed GCs of \dg{} as well as local GCs \citep{Harris1996, Georgiev2009}. We select sources between $m_{F606W} = 19.5$ mag and $m_{F606W}$ = 27.1 mag\footnote{This is based on the depths of our images (Sec.~\ref{sec:DataDepth}). However, in App.~ \ref{app:completeness}, we estimate our completeness based on injecting mock GCs in the images, fully capturing the entire process of source detection. We decide, conservatively, to use the fainter value for the GC pre-selection (27.1 mag) and, thereafter, correct the number estimates using the completeness tests.}. The faint limit corresponds to the $5\sigma$ limiting depth calculated in Sec. \ref{sec:DataDepth}. The bright limit corresponds to the magnitude of the the brightest GC from the MW and from nearby dwarfs \citep{Harris1996, Georgiev2009} in the case that \dg{} is at 20.7 Mpc (this is a conservative limit, since a GC this bright would be fainter if \dg{} is at 26.5 Mpc; see Appendix ~\ref{app:comparisongcs} for a comparison between the MW and nearby dwarfs GCs and the GCs of \dg). 

The effective radius and magnitude are taken from our deepest image (WFC3/\textit{F606W}) while the ellipticity used is the average between the ellipticity from the three HST bands in order to maximise the reliability of the measurement. The ellipticity and effective radius are measured using \texttt{SExtractor}. To obtain an approximate value for the $R_e$, we analytically deconvolve the FWHM of the candidate GC obtained from \sextractor\ with the FWHM of the PSF of the images under a Gaussian assumption. To convert to a physical size, we assume a distance of 20.7 Mpc.\footnote{Assuming a distance of 26.5 Mpc, the final catalogue has two candidates less.}

The pre-selection criteria is the following:
\begin{itemize}
  \item 0 pc $\leq$ $R_{e}$ $\leq$ 10 pc
  \item 0 $\leq$ ellipticity $\leq$ 0.2
  \item 19.5 mag $\leq$ $m_{F606W}$ $\leq$ 27.1 mag 
\end{itemize}

Fig. \ref{fig:ReEllipMagHists} shows the histograms of $R_e$, ellipticity and $m_{F606W}$ of all the sources detected with \sextractor{} (grey), the confirmed GCs in \citet{Muller2020} and \citet{Haacke2025} (hashed red) and the GC candidates based in our preliminary selection (blue). A total of 128 objects, including the 20 spectroscopically confirmed GCs, fulfil the criteria above.

\subsection{Colour-colour selection}\label{sec:colour-colour}
\begin{figure}
    \centering
    \includegraphics[width=0.85\columnwidth]{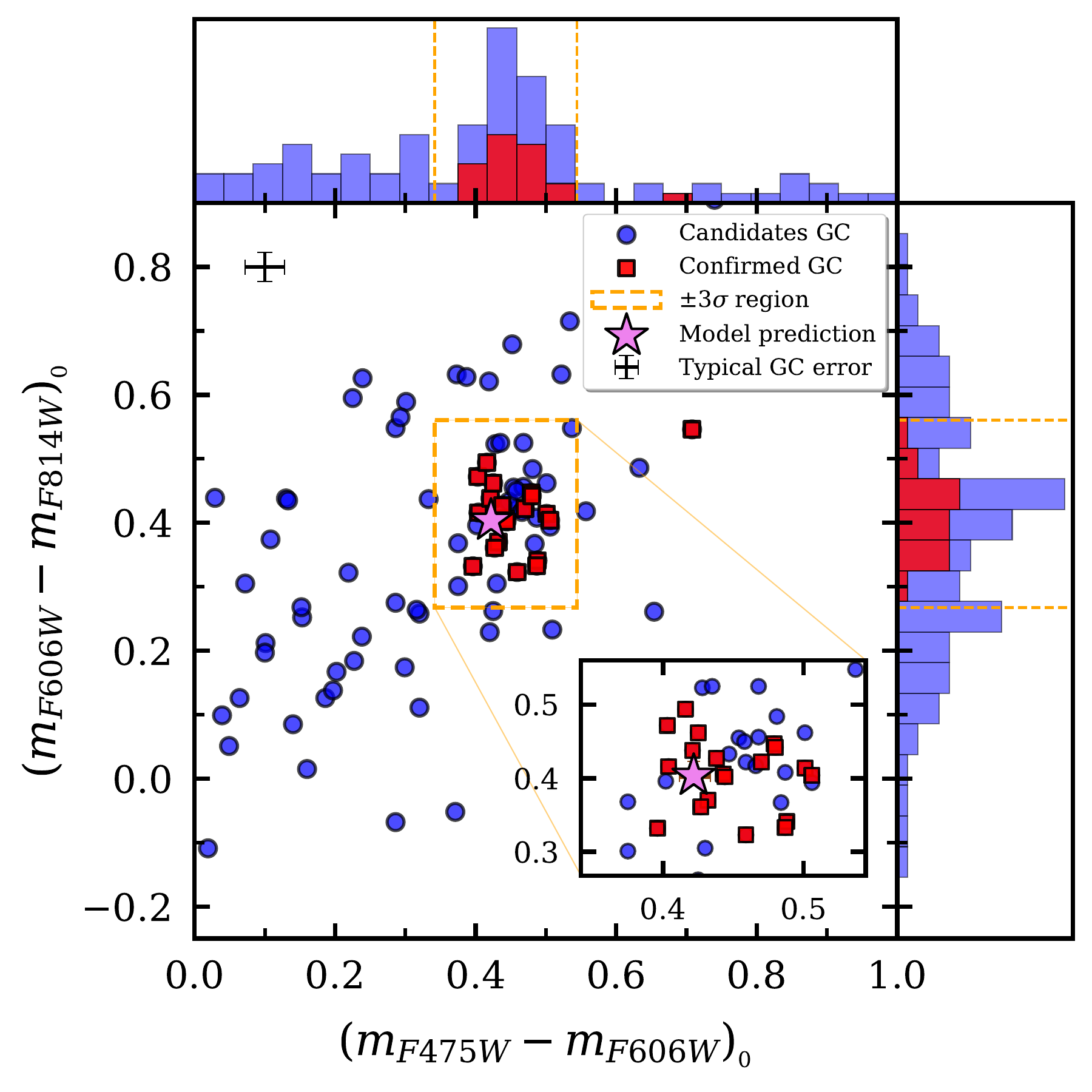}
    \caption{The $(m_{F475W} - m_{F606W})_0$ vs $(m_{F606W} - m_{F814W})_0$ colour-colour diagram of the initial sample of GC candidates. The 20 spectroscopically confirmed GCs \citep{Muller2020, Haacke2025} are shown as red squares. The orange dashed box indicates the region we have selected for our next sample of GC candidates and corresponds to $\pm3\sigma$ around the median colours of the confirmed GCs (excluding the anomalous object, see Sec. \ref{sec:colour-colour}). The violet star, and its error bar, shows the prediction from \citet{Bruzual2003} models for an SSP of [Fe/H] = -1.44 dex and 9.1 Gyr \citep{Muller2020}. The inset shows a zoom-in into the dashed orange box for ease of viewing. The error bars shown in the upper-left corner are the errors of an object located at the peak of the GCLF ($m_{F606W}$ = 23.77 mag, assuming a distance of 20.7 Mpc to the galaxy).}
    \label{fig:colourColourDiagram}
\end{figure}

Once we have our GC candidates, the next step is to narrow down the selection based on the colours of the GCs. Fig. \ref{fig:colourColourDiagram} shows the  $(m_{F475W} - m_{F606W})_0$ vs. $(m_{F606W} - m_{F814W})_0$ colour-colour diagram. The 20 spectroscopically confirmed GCs in \citet{Muller2020} and \citet{Haacke2025} are indicated as red squares. All the spectroscopically confirmed GCs but one occupy a small region in the colour-colour space. This anomalous object was already pointed out by \citet{Haacke2025} as an object with a remarkable different colour. This object has a $(m_{F475W} - m_{F606W})_0$ colour at  $\sim7.8\sigma$ and a $(m_{F606W} - m_{F814W})_0$ colour at $\sim2.7\sigma$ from the rest of the GCs. In light of this anomaly, we assume that this object is intrinsically different from the rest of the GC system (perhaps an intragroup GC) and will not be considered as a bona fide GC for the rest of the analysis. In Appendix \ref{app:anomalous} we explore how the results change if we include it in the analysis. 

We select objects with colours between $0.34 < (m_{F475W} - m_{F606W})_0 < 0.54$ mag and, simultaneously, $0.26 < (m_{F606W} - m_{F814W})_0 < 0.55$ mag (orange box). Similar to \citet{Montes2020, Montes2021}, this selection region is based on the $\pm 3\sigma$ around the median colours of the 20 confirmed GCs (i.e. ($m_{F475W} - m_{F606W})_0 \sim 0.44$ and ($m_{F606W} - m_{F814W})_0 \sim 0.41$), with $\sigma$ being their colour dispersion ($\sigma_{(m_{F475W} - m_{F606W})_0} \sim 0.034$ and $\sigma_{(m_{F606W} - m_{F814W})_0} \sim 0.049$). The violet star in Fig.~\ref{fig:colourColourDiagram} shows the colour predictions using the simple stellar population models from \citet{Bruzual2003} based on the metallicities ([Fe/H] = $-1.44^{+0.10}_{-0.07}$ dex) and ages ($9.1^{+3.0}_{-0.8}$ Gyr) of the confirmed GCs of \dg{} obtained in \citet{Muller2020}. The colours of the model are: $(m_{F475W} - m_{F606W}) = 0.42^{+0.012}_{-0.010}$ and $(m_{F606W} - m_{F814W}) = 0.40^{+0.020}_{-0.013}$, in agreement with our selection region. After the colour-colour selection, only 38 potential GC candidates match our criteria. 

\subsubsection*{OSIRIS+ \textit{u}-band photometry} \label{u-band_phot}

We take advantage of the availability of the $u$-band from OSIRIS+ to further refine our GC candidate sample. The $u$-band is especially useful to separate GCs from other sources such as foreground stars, as horizontal branch stars present in these metal-poor GCs contribute to its flux \citep[e.g.][]{Taylor2017}. As mentioned in Sec.~\ref{sec:DataDepth}, this colour criterion is applied separately because of the different properties of the GTC and HST data. 

The photometry in the $u$-band data is performed at the locations where the sources were detected in the HST images. The apertures used are based on the PSF of the OSIRIS+ image, modelled in the same way as the PSF in the HST bands (Sec \ref{sec:psfImages}). Thus, the diameter used for the aperture is $2\times$FWHM of the PSF ($D = 2\farcs2 = 8.6$ pix). Analogous corrections to those explained in Sec. \ref{sec:AperAndCorrect} have been calculated for the $u$-band. Thus, we obtain that a correction of $-0.4$ mag is needed for the first step (from the aperture to the end of the profile), and a second correction of $-0.04$ mag when extrapolating to infinity. Again, these are applied as $m_{corr} = m_{ini} + m_{corr1} + m_{corr2}$. Afterwards, the photometry has also been corrected from extinction ($A_{u} = 0.269$ mag).

\begin{figure}
    \centering
    \includegraphics[width=1\linewidth]{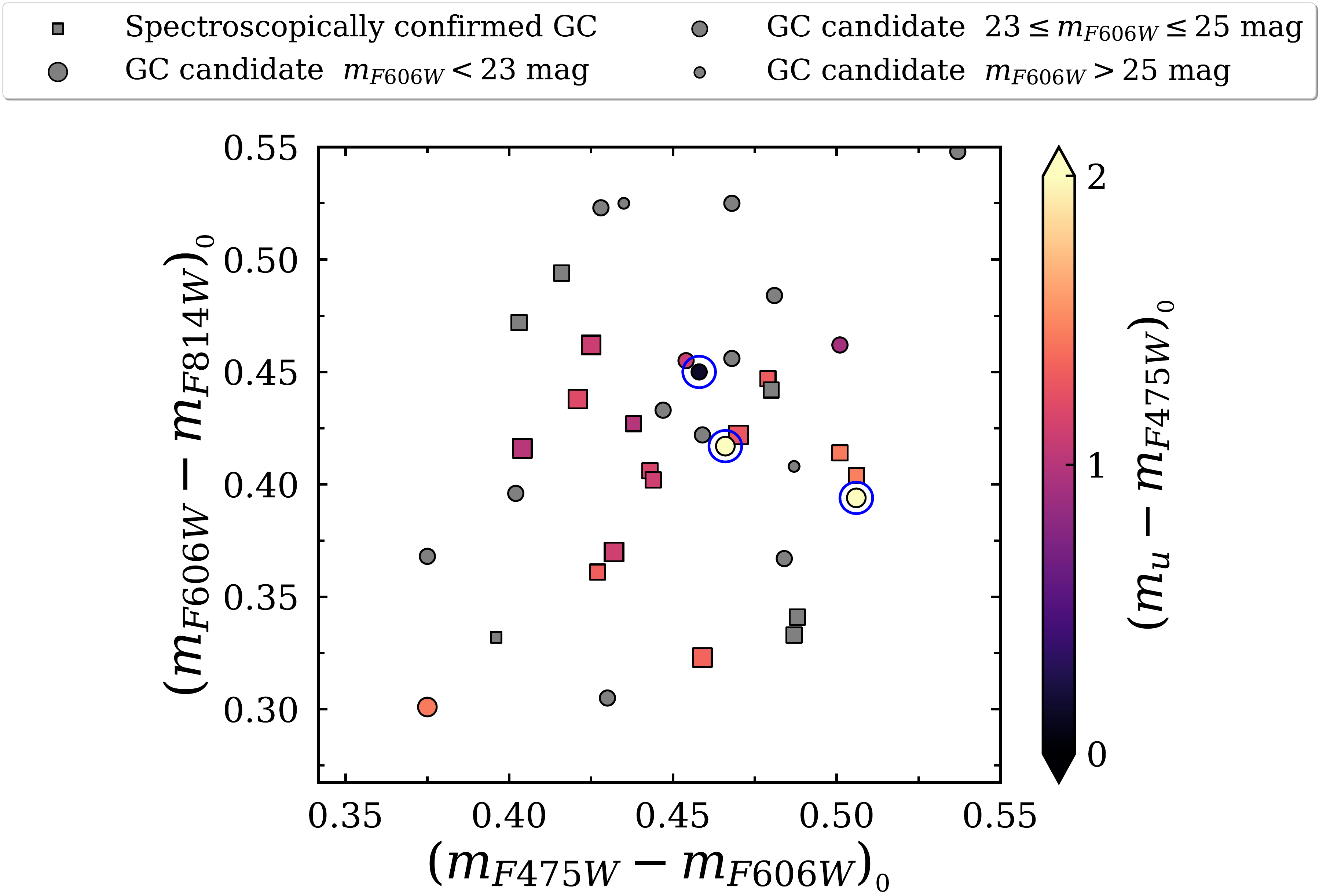}
    \caption{The $(m_{F475W} - m_{F606W})_0$ vs $(m_{F606W} - m_{F814W})_0$ colour-colour diagram of GC candidates, colour coded with the $(m_u-m_{F475W})_0$ colour. The 20 spectroscopically confirmed GCs \citep{Muller2020, Haacke2025} are shown as squares and the candidates as circles. The size of the marker indicates the $m_{F606W}$ of the source, as shown in the legend. The sources in grey are objects that are too faint to be measured in the $u$-band. The blue open circles highlight those objects rejected based on their $(m_u-m_{F475W})_0$ colour.}
    \label{fig:colourDiagram_u}
\end{figure}

\begin{figure}
    \centering
    \includegraphics[width=0.65\linewidth]{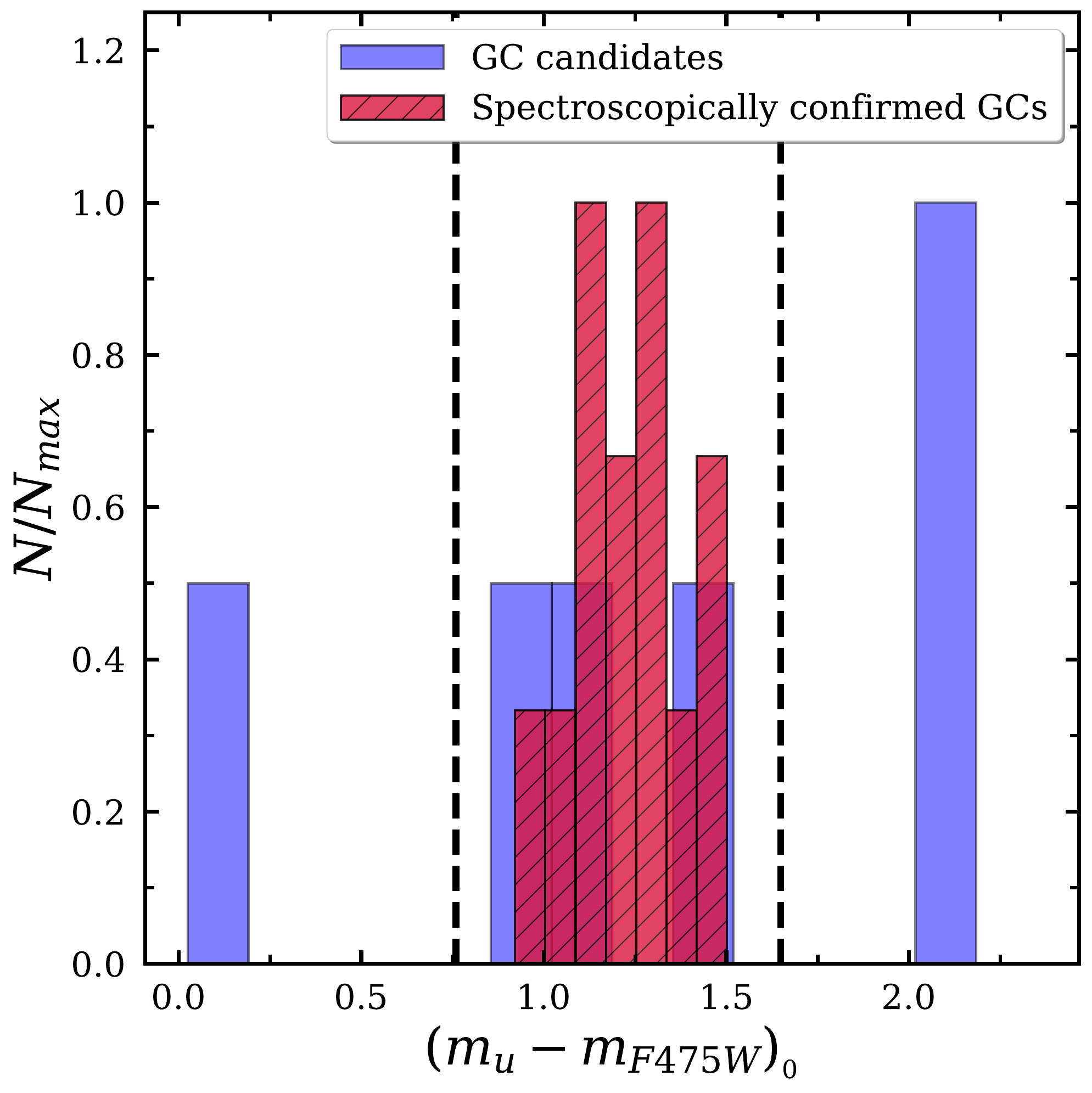}
    \caption{Normalised histograms of the $(m_u-m_{F475W})_0$ colour of the GCs of MATLAS2019; N is the number of sources in each bin and $N_{max}$ the maximum of their respective histograms. The red hatched histogram shows the spectroscopically confirmed GCs and the blue histogram the GC candidates. Only sources with trustable magnitude in the $u-$band are shown (i.e. $m_u$ < 25.8, see Sec. \ref{u-band_phot}). The dashed vertical black lines indicate the $\pm3\sigma$ cut applied for GC selection.}
    \label{fig:u-475Hist}
\end{figure}

Fig.~\ref{fig:colourDiagram_u} shows the same colour-colour diagram as in Fig.~\ref{fig:colourColourDiagram}, but in this case the GC candidates are colour-coded by the $(m_u-m_{F475W})_0$ colour. We obtain a median $(m_u-m_{F475W})_0$ colour of 1.20 mag and a standard deviation of 0.15 mag for the spectroscopically confirmed GCs. The expected $(m_u-m_{F475W})_0$ colour based on the properties of the spectroscopically confirmed GCs and on the \citet{Bruzual2003} models is $1.16^{+0.06}_{-0.02}$. We select those objects within a range of $\pm 3\sigma$ (i.e. $0.75 < (m_u-m_{F475W})_0 < 1.65$). Note that we can only apply this selection to objects that are observed in the $u$-band, i.e. m$_u$< 25.8 mag (see Appendix \ref{app:completeness}). Applying this colour selection, we end up with 35 GC candidates. 
We highlight the GCs rejected by including the $u$-band in Fig. \ref{fig:colourDiagram_u}. Fig. \ref{fig:u-475Hist} shows the distribution of the $(m_u-m_{F475W})_0$ colours for the candidates and for the spectroscopically confirmed GCs.

To summarize, the colour ranges used in this work are:

\begin{itemize}
  \item 0.34 mag $<$ $(m_{F475W} - m_{F606W})_0$ $<$ 0.54 mag
  \item 0.26 mag $<$ $(m_{F606W} - m_{F814W})_0$ $<$ 0.55 mag
  \item 0.75 mag $<$ $(m_u-m_{F475W})_0$ $<$ 1.65 mag
\end{itemize}

\begin{figure*}[h]
    \centering
    \includegraphics[width=13cm]{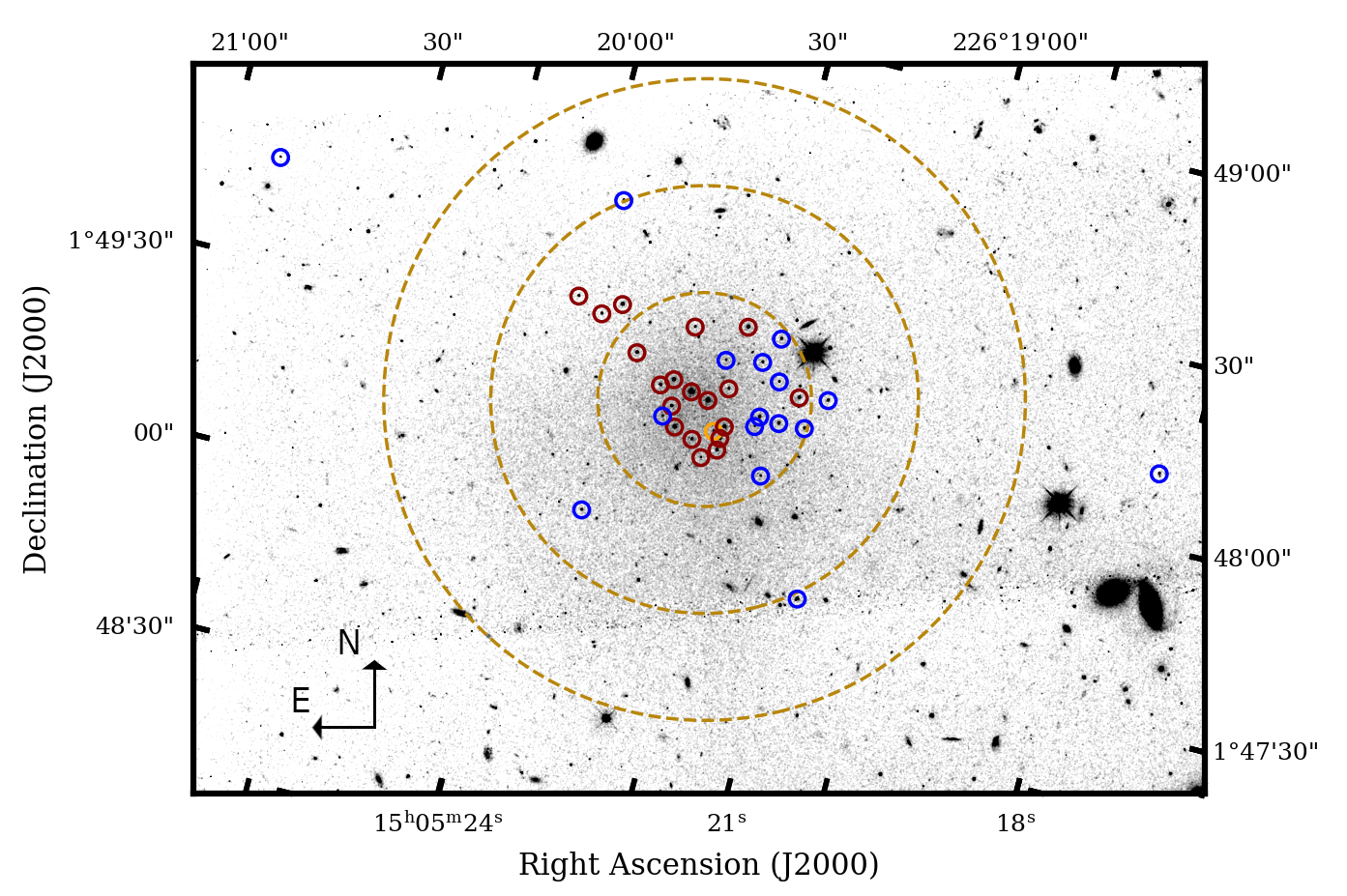}
    \caption{HST WFC3/\textit{F606W} image convolved with a Gaussian kernel ($\sigma=1.5$) showing the spatial distribution of the GC candidates identified in this work.  The red circles indicate the GCs spectroscopically confirmed in \cite{Muller2020} and \cite{Haacke2025} while the blue circles highlight the GCs candidates and the orange circle represent the anomalous GC. The yellow dashed circles indicate 1$R_{e}$, 2$R_{e}$ and 3$R_{e}$ of the galaxy ($R_e =$ \matlasEffRad, \citealt{Muller2020}).}
    \label{fig:imageDistribution}
\end{figure*}

The coordinates and parameters measured of all the objects are compiled in Table \ref{tab:gcPopulation} and Fig. \ref{fig:imageDistribution} shows a smoothed WFC3/\textit{F606W} \dg{} image with the GC candidates marked as open blue circles.

\subsection{Final number of GC candidates} \label{sec:finalnumber}

Errors in photometry, ellipticity, and $R_e$ could affect our identification of GCs. To determine the completeness and characterize the errors on the photometry and the morphological parameters derived above, we perform a series of tests by injecting artificial GCs on the images. Further details on these tests are given in Appendix \ref{app:completeness}.

We use the results of these tests to assess how the errors in colour, magnitude, ellipticity, and effective radius impact the selection of GC candidates. To do so, we performed 10000 realisations where the parameters of the detected sources (i.e. detections matched among the three HST filters, excluding the spectroscopically confirmed GCs) have been perturbed by drawing a value from the distribution of their errors, assuming a Gaussian distribution. The aim is to statistically characterise the identification of GC candidates of \dg{} based on the previously described procedure. For each of these realisations, we derive a candidate GC sample. The median number of detected GCs is 32, with a standard deviation of 3. Consequently, although we identify 35 sources as GC candidates, the expected population of GCs of the galaxy is $32\pm3$.

We also estimate how many GCs might be missing due to the limited depth of our images. In dwarf galaxies, the GC luminosity function (GCLF) peaks around $M_V$(Vega) $= -7.3$ with a width of $\sigma\sim1$  \citep[e.g.][]{Miller2007}, although other studies suggest narrower distribution $\sigma\sim0.6$ (e.g. \citealt{Carlsten2022}). Thus, the expected GCLF would peak at $m_{F606W} (AB) = 24.13$ mag\footnote{We use the conversion $V_{606}(AB)$ = $V$(Vega) $-\,0.145$ mag. This is derived computing $V_{606}(AB)$ and $V$(Vega) of a simple stellar population model from \citet{Vazdekis2016} with [Fe/H] = $-1.44$ and age 10 Gyr.} if the galaxy is at 20.7 Mpc or $m_{F606W} (AB) = 24.67$ mag if it is at 26.5 Mpc. Being conservative and assuming a wider GCLF in dwarfs (i.e. 1 mag), the $2\sigma$ value from the peak will be located at $m_{F606W}$ = 26.13 mag at 20.7 Mpc or $m_{F606W}$ = 26.67 mag at 26.5 Mpc. Based on completeness tests (Appendix \ref{app:completeness}), we reach 90\% completeness in WFC3/\textit{F606W} around 26.5 mag, this means that we are only missing $\sim2.5\%$ of the sources, regardless of the assumed distance. This adds 1 GC to our statistical estimation.

All in all, we end up with a population of 35 GCs for \dg. After characterising and exploring the errors in the selection parameters, the expected number of GCs ends up being $32\pm3$ objects. Including the completeness correction due to the depth of our images, the final population is \finalPopulation.

\subsection{Contamination}
\label{sec:Contamination}

Due to the slightly different fields covered by the different images used in this work, it is not straightforward to define a control field for assessing the contamination. We have estimated it by assuming as our control field the region outside 2$R_e$ \citep[$R_e$ = \matlasEffRad, ][]{Muller2020} of the galaxy (being conservative with respect to \citealt{Muller2021}, where it is found that the GC radial distribution falls to background levels beyond 1.75$R_e$), where 4 candidates are found. Doing this, we find that the contamination level in our sample is $\sim1$ GC. As expected ---due to the exhaustive morphological and colour criteria--- the contamination is low. Moreover, two of the four candidates that lie outside $2R_e$ have GC-like profiles, in contradiction with what we would expect from contaminants. Thus, we judge sensible to be conservative and not to consider these objects as interlopers in spite of their far location to the centre of \dg.

\begin{figure}
    \centering    
  \includegraphics[width=0.8\linewidth,trim=5 35 5 40,clip]{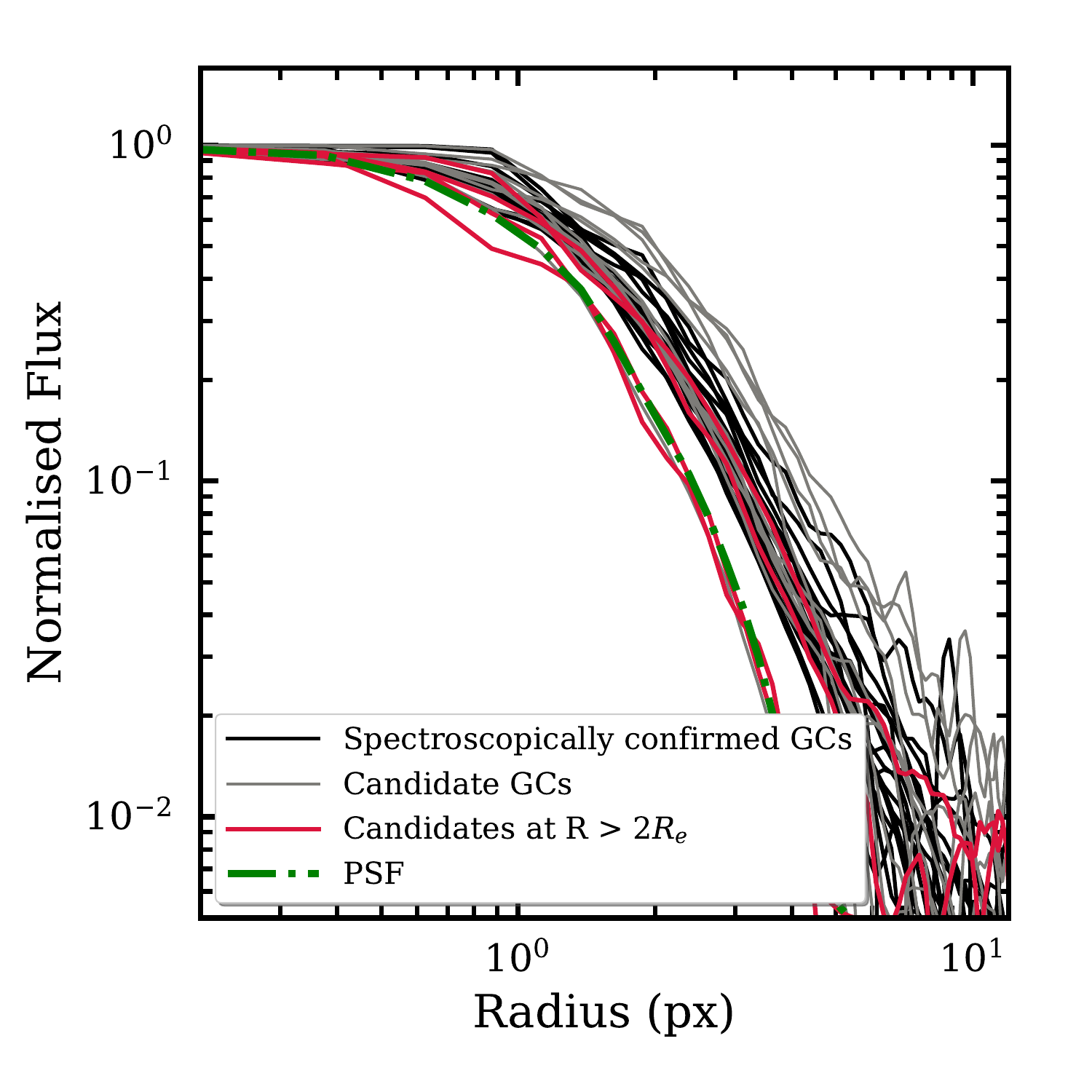}
    \caption{Comparison of radial profiles in the WFC3/\textit{F606W} band. We show the profiles of the spectroscopically confirmed GCs, the GC candidates obtained in this study in grey, the profile of the PSF from stars in green, and the profiles of 4 candidates located at more that 2$R_e$ from the galaxy's centre in red.}
    \label{fig:strangeProfiles}
\end{figure}

Fig.~\ref{fig:strangeProfiles} shows the radial profiles for the GC candidates in this work. The confirmed GCs from \cite{Muller2020} and \cite{Haacke2025} are in black, the new candidates identified in this work in grey, and the PSF is shown as the dashed green line. The candidates located $>2R_e$ from the centre of \dg{} are shown in red. Two of them have profiles that are similar to the profiles of confirmed GCs, suggesting that they are likely GCs. On the other hand, two of them deviate from the general trend and resemble a point source. These point-like sources are strong interlopers candidates, but we have not discarded them as more information is needed. In Table \ref{tab:gcPopulation} these candidates are flagged, along with another GC candidate with also a point-like profile. 

\section{Stellar properties of \dg}\label{sec:optical}

The ultra-deep imaging of \dg{} with OSIRIS+ allows us to assess its optical morphology in detail for the first time, as previous imaging \citep[e.g.][]{Forbes2020, Habas2020, Muller2021, Danieli2022} was not deep enough to reveal the low surface brightness features of the galaxy (i.e. mainly tidal features or asymmetries in the outskirts). The outer parts of low-mass galaxies, such as \dg{}, can provide direct evidence for their hierarchical merging assembly. For example, asymmetries or other perturbations indicate tidal interactions with nearby galaxies \citep{Johnston_2002, Montes2020, Golini2024}. In this section, we explore the deep images from OSIRIS+ to gain a deeper understanding of the physical processes that this galaxy may have undergone.

\subsection{Surface brightness radial profiles}\label{sec:sbprofs}

\begin{figure*}
\centering
    \includegraphics[trim=3 3 3 3,clip,width=0.7\textwidth]{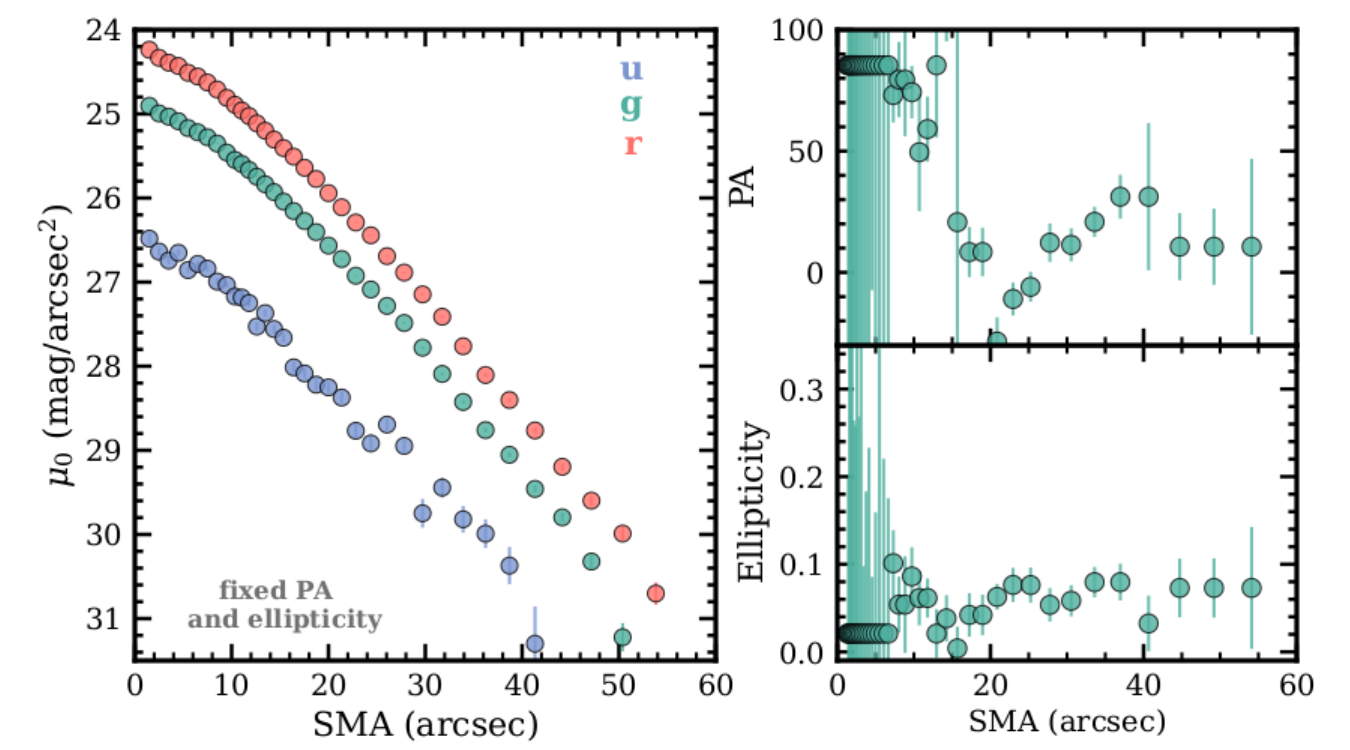}
    \caption{Surface brightness, ellipticity and PA radial profiles of \dg. The left panel shows the $u$-band (blue), $g$-band (teal) and $r$-band (red) surface brightness profiles as a function of the semi-major axis obtained with a fixed centre and ellipticity ($\epsilon$ = 0.07). The surface brightness profiles have been corrected from extinction. The right panel shows the position angle (top) and ellipticity (bottom) profiles from letting the parameters of the fit vary freely.}
    \label{fig:ellipsefitting}
\end{figure*}

Signs of interactions will show up in the radial surface brightness profiles as an excess of light at large radii and deviations from the morphology of the inner parts of the galaxy (see e.g. \citealt{Johnston_2002}). To derive the surface brightness profiles of \dg{} we follow the steps below.

We run \texttt{SExtractor} in a deep combined $g+r$ image and obtain the segmentation map to mask background and foreground sources. This mask was visually inspected to manually mask any remaining light that was missed by \texttt{SExtractor}. Once all sources of contamination are masked out, we run \texttt{ellipse} \citep{Jedrzejewski1987} in \texttt{photutils} \citep{larrybradley2023} leaving all the parameters free to derive the radial profiles of the galaxy's ellipticity and position angle (PA). 

The right panel of Fig. \ref{fig:ellipsefitting} shows the PA (top) and ellipticity (bottom) radial profiles of \dg. In the inner parts of the galaxy ($R<30$\arcsec) we see a large scatter in PA and ellipticity due to the conservative masking caused by the presence of bright GCs. In the outer parts, however, both ellipticity and PA appear to be constant, indicating that the galaxy is quite symmetrical (see Fig.~\ref{fig:rgbgrey}). The median values are $\rm{PA} = 10\fdg5$, and $b/a = 0.07$, at $R>30$\arcsec. We also obtained the centre of the galaxy (RA= $15^h$$05^m$$20.17^s$, Dec= $+1^d$$48^m$$44.91^s$), from the output of \texttt{ellipse}.

To derive the radial profiles of \dg, in the $u$, $g$ and $r$ bands, we use a similar approach as in \citet{Montes2021}. We extracted the radial profiles in elliptical bins at different radial distances from the centre of the galaxy, fixing the ellipticity and PA of the bin to the average values obtained above. We used a custom python code to derive the profiles. For each of the radial bins, the intensity was obtained as the $3\sigma$-clipped median of the pixel values. The errors are calculated as a combination of the Poisson noise in each annulus and the error given by the distribution of background pixels in each image. The profiles are corrected for the extinction of the Galaxy (E(B-V) = 0.046, see also Sec.~\ref{sec:GCSelection}). The left panel of Fig.~\ref{fig:ellipsefitting} shows the surface brightness profiles of \dg{} in the $u$ (blue), $g$ (teal) and $r$ (red) bands. From the light profiles we derive an effective radius of $R_e = 16.\!^{\prime\prime}0 \pm 0.5$, compatible with the value given by \cite{Muller2020} ($17.\!^{\prime\prime}2\pm0.2)$.

\subsection{Colour and stellar mass density profiles}

\begin{figure}
\centering
    \includegraphics[width=0.35\textwidth]{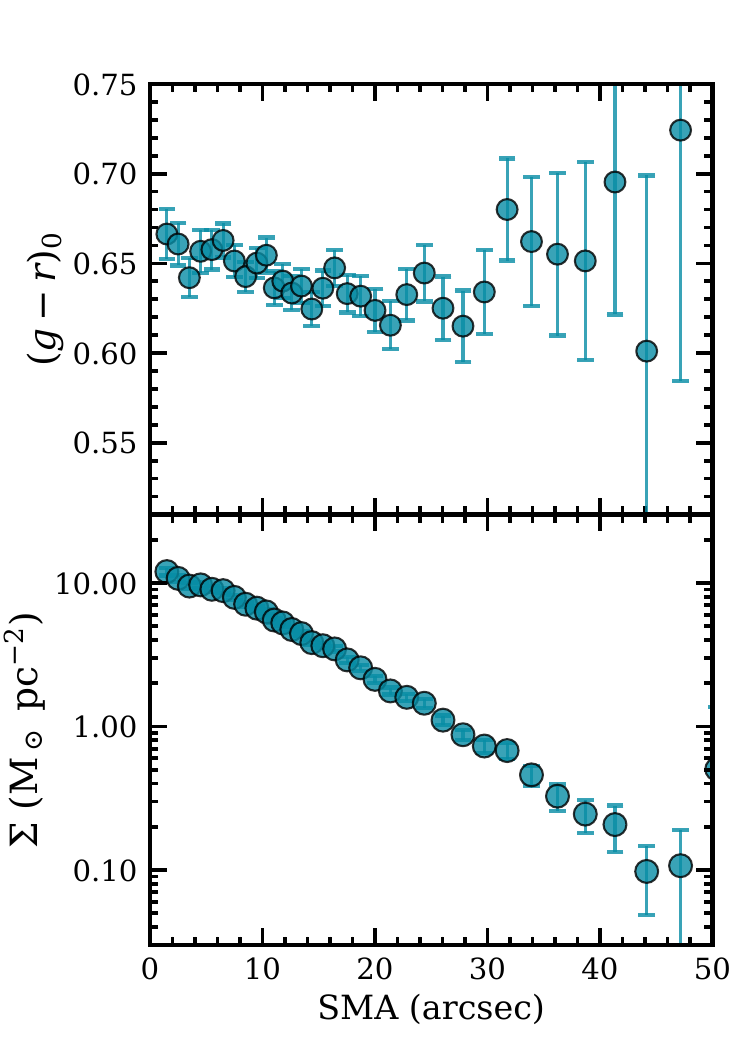}
    \caption{Upper panel: OSIRIS+ colour $(g-r)_0$ radial profile of \dg. Bottom panel: The surface stellar mass density profile of \dg. The density profile shows a break around $\sim33$\arcsec, coinciding with a change in the color profile. }
    \label{fig:profiles}
\end{figure}

Once we derived the surface brightness profiles, we can study the radial variations of the galaxy's stellar populations. These profiles are used to derive the $(g-r)_0$ colour profile of the galaxy, shown in the top panel of Fig.~\ref{fig:profiles}. The $(g-r)_0$ profile appears quite flat, with an average colour of $\sim0.64$ down to R$<30$\arcsec. For $R>30\arcsec$, the colour raises slightly up to a value of $\sim0.7$. This ``u-shape'' profiles have been found very often in previous works as indicative of the end of in-situ star formation (see e.g. \citealt{Azzollini2008, Bakos2008}).

The bottom panel in Fig. \ref{fig:profiles} shows the stellar mass surface density of \dg. This profile was derived using the prescriptions given in \citet{Bakos2008} that linked the surface brightness profiles in the $g$-band to radial variation of the mass-to-light ratio. The mass-to-light ratio was obtained from \citet{Roediger2015}, assuming a \citet{Chabrier2003} IMF and using the $(g-r)_0$ colour profile of the galaxy. 

From the stellar mass surface density profile, we derived the total stellar mass of \dg{} from the mass density profile, assuming circular symmetry, is $M_* = 9.2^{+3.0}_{-2.3} \times 10^7$ M$_\odot$ at a distance of $20.7^{+2.3}_{-2.1}$ \citep{Muller2021}, or $1.5 \pm 0.2 \times 10^8$ M$_\odot$ if the galaxy is located at $26.5 \pm 0.8$ Mpc \citep{Danieli2022}.

The stellar mass surface density profile of \dg{} shows an exponential decrease up to $\sim33$\arcsec, corresponding to 3.3 (4.2) kpc assuming a distance of 20.7 (26.5) Mpc to the galaxy. At this radius there is a change in the slope, or break, of the density profile that coincides with a change in the colour profile of the galaxy. This radius also coincides with the visual extension of the galaxy (see Fig. \ref{fig:colourImage}). This break is similar to those found for dwarfs of the same stellar mass in \cite{Chamba2022}, occurring at a similar stellar mass density ($\Sigma_* \sim 0.5$ M$_\odot$/pc$^2$). The presence of this truncation signals the edge of the galaxy, beyond which the galaxy could not reach the critical gas density threshold for star formation \citep{Trujillo2020}. 

\section{Discussion}

In this paper we undertake a comprehensive identification of GC candidates for \dg{}, in order to shed light on the current disagreement about this GC population. The candidates have been identified using photometry obtained from archival HST and GTC deep data, and the stellar properties of the galaxy have also been analysed. Here we discuss the implications of the GC counts found in this galaxy and how they relate to the current debate.

\subsection{Comparison with previous studies} \label{sec:comparisonWithStudies}

There is disagreement about the total number of GCs present in \dg. \citet{Forbes2020b} estimated a total of $\sim$45 GCs in this galaxy, based on \citet{Forbes2019} where about 20 compact sources were identified within 2$R_e$ of the galaxy in ground-based broadband imaging, and assuming a typical GCLF to correct for completeness. Given the differences in the datasets, mainly resolution, we cannot compare with their results.

\citet{Muller2021} used HST data to estimate a total number of GCs for \dg{} of $26 \pm 6$ GCs. They used a threshold in source radius, ellipticity, and $(m_{F606W} - m_{F814W})_0$ colour range finding an initial sample of 49 GC candidates within $1.75R_e$, which is narrowed down to $26 \pm 6$ using Bayesian considerations. They also estimated the population inside $1.75R_e$ by taking into account the expected contamination in a reference background field, resulting in a population of $36 \pm 6$. We find that our estimate for the total population of GCs (\finalPopulation) is compatible with both estimations. By using three filters information and limiting the $R_e$ of the sources, our uncertainty estimates are lower than in \citet{Muller2021} ($\pm6$ vs $\pm3$). If we only use the same filters (\textit{F606W} and \textit{F814W}) but with our selection criteria, we find a population of $34\pm2$ inside $1.75R_e$, which is still consistent with their results in the same region and with the same filters. 

\citet{Danieli2022} also used HST data to explore the GC population of \dg. In their case, they found a total $54 \pm 9$ GCs ($50.4$ within 2$R_e$, before completeness corrections). This would imply that \dg{} is one of the UDGs with the richest GC system. \citet{Danieli2022} selection is based in the FWHM and colour criteria from the spectroscopically confirmed GCs of the galaxy, similar to this work. However, the results are not compatible. Using only the same filters as \citet{Danieli2022} (\textit{F606W} and \textit{F475W}), but applying our criteria, we find $31\pm2$ candidates within 2$R_e$, as opposed to the 50.4 GCs they find in the same area. 

\citet{Danieli2022} do not apply any morphology criteria when the GC candidates have magnitudes between $m_{F606W} = 25$ mag and  $m_{F606W} = 26.5$, which might contribute to the difference in GC numbers that we find. Nevertheless, the main difference between both studies is the range of colours used. \citet{Danieli2022} uses different colour criteria depending on the magnitude of the source, with colour bin widths ranging from 0.4 mag to 0.72 mag. On the other hand, we use a colour range of $\sim 0.2$ mag for $(m_{F475W}-m_{F606W})_0$ and of $\sim 0.3$ for $(m_{F606W}-m_{F814W})_0$ (which is $\pm3\sigma$ the median colour of the confirmed GCs). Defining the acceptance region is reaching a compromise between the accepted contamination level and the ability to correctly identify the properties of the faintest sources. Based on the photometric errors of the data (see Appendix \ref{app:completeness}) and on the expected GCLF, the faintest analysed objects will have $m_{F606W} \sim 26.5$. This implies that the magnitude errors for the faintest candidates are around 0.15 mag. We consider that a selection region of around two times this value is reasonable enough and will minimise contaminants. Nevertheless, as a sanity check, applying a wider colour range (e.g. $0.2 < (m_{F475W} - m_{F606W})_0 < 0.8$), we obtain $43\pm2$ GCs inside $2R_e$, more in line with \citet{Danieli2022}. Our narrower colour range, combined with the addition of two extra bands and taking into account the ellipticity of the objects to separate the GCs from background sources, produces the difference. 

A recent study of the GC system of MATLAS-2019 \citep{Marleau2024}, estimated a population of $38\pm7$ GCs. The data used in this study combined the data used by \cite{Muller2021} with an extra set of observations (GO-16257 and GO-16711; PI: Marleau) taken with the same instrument and filters (HST/ACS, filters \textit{F606W} and \textit{F814W}). The parameters used to select candidates were colour, ellipticity, $R_e$ and concentration index. Our result is lower but compatible with \cite{Marleau2024}. As in the case of \cite{Danieli2022}, the higher number of GCs 
is mainly due to the colour range accepted. Our selection window in $(m_{F606W} - m_{F814W})_0$ is narrower (being from 0.5 to 1.2 in \citealt{Marleau2024}), but also we filter GCs using two extra colours, $(m_{F475W} - m_{F606W})_0$ and $(m_u-m_{F475W})_0$, which allows us to further reject contaminants in our sample. 
Note that, even though we do not explicitly use the concentration index in this work, we are including that information when we analyse the light profiles of the candidates.

The fraction of light which corresponds to the GCs in \dg{} in our study is roughly $ 5.4\pm0.5\%$ (V-band, assuming  $M_{V}(Vega) = -7.66$\footnote{Using $m_{AB}-m_{Vega} = 0.02$ from \cite{Blanton2007}.} for each globular cluster and $M_{V}(AB) = -14.6$ mag for the galaxy from \citealt{Danieli2022}\footnote{We obtain $8.9\pm2.4\%$ assuming $M_{V} = -14.1\pm0.2$ mag from \cite{Muller2021}.}), in contrast to the larger contribution ($12.9\% \pm 0.6\%$ in the V-band) proposed by \citet{Danieli2022}. As expected in an UDG, the fraction found here is larger than the ratios of normal galaxies (e.g,. MW has $\sim0.1\%$, assuming 160 GC and $M_V = -20.9$, \citealt{Bergh1999}) but in line with other UDGs and dwarfs \citep[e.g.][]{Forbes2020}. From the results obtained in this work, the specific frequency of \dg{} (defined as $S_N = N_{GC}\times 10^{0.4(M_V+15)}$), is $78 \pm 21$ (using $M_V = -14.1 \pm 0.2$ mag from \citealt{Muller2021}). This value is consistent with those reported in \citet{Muller2021} and \citet{Marleau2024} ($59\pm14$ and $90.97\pm16.49$, respectively).

\subsection{Spatial distribution of the GCs} \label{sec:SpatialDistributionOfGC}

In this section, we discuss the spatial distribution of the GCs in \dg{} in order to obtain clues about the nature of this galaxy.
Note that while we statistically estimate a population of \finalPopulation{} (Sec. \ref{sec:finalnumber}), the actual number of GCs that we identify is 35. Since based on the statistical analysis we cannot identify which specific sources are interlopers, we will use the 35 identified candidates for the study of the spatial distribution of the GC system.

\begin{figure}[htb]
    \centering
    \includegraphics[width=8cm]{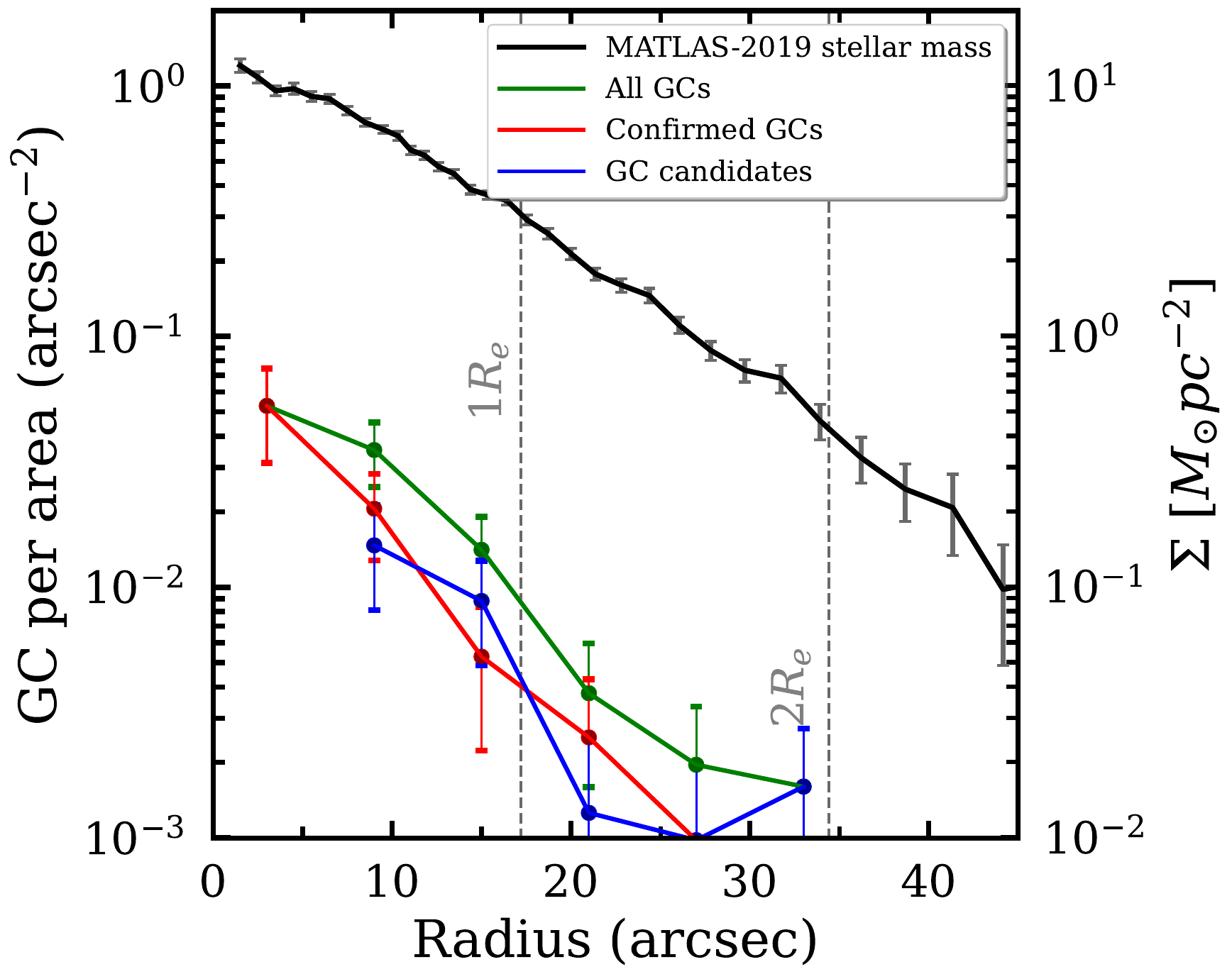}
    \caption{Radial distribution of GC candidates in \dg{}. The red profile represents the spectroscopically confirmed GCs, the GC candidates obtained in this study are in blue, and the whole GC system in green. The black line is the mass density profile of \dg{}, as indicated in the right Y-axis. The errors in the GC profiles are the Poissonian error of the number of GCs per area in each annulus (i.e. $\sqrt{N_{GC}} / A$). The dashed vertical grey lines indicate 1$R_{e}$ and 2$R_{e}$ of the galaxy ($R_{e} =$ \matlasEffRad, \citealt{Muller2020})}
    \label{fig:radialProf}
\end{figure}

Fig.~\ref{fig:imageDistribution} shows that most of the identified GC candidates are inside 1$R_e$ (26; i.e. around $80\%$ of the total GC number) and 2$R_e$ (31), while 4 candidates are beyond 2 $R_e$ \citep[$R_e =$ \matlasEffRad ,][]{Muller2020}. To measure the number density profile of the GC candidates, we measure the number of GCs in a set of annuli up to 45\arcsec (about $2.5R_e$ of the galaxy). The width of the annuli is fixed at 6\arcsec, which is about $\frac{1}{3}R_e$. The resulting profiles for the different samples (confirmed in red, candidates in blue, and the whole GC system in green) are shown in Fig.~\ref{fig:radialProf}. For comparison, we also show the stellar mass density profile of \dg{} in black. To convert GC density (in arcsec) to mass density (in pc) we assume a mass of $10^5$ M$_\odot$ for each GC and a distance of 20.7 Mpc \citep{Muller2021} to \dg. Notably, there is a lack of GC candidates towards the centre of the galaxy. This is because there are more bright GCs in this region, that can be more easily observed with spectroscopy, and the spectroscopic coverage is also better. 

The GC half-number radius (i.e. the radius containing half of the GCs, $R_{e,GC}$) is $12.0 \pm 0.1$ arcsec. Comparing these values with the $R_e$ of the galaxy (\matlasEffRad), we find a ratio of $R_{e,GC}/R_{e} = 0.7$. This indicates that the GC system is less extended than the stellar light component of the galaxy, also seen in Fig.~\ref{fig:radialProf}. This ratio is remarkably smaller from what other authors have claimed for other UDGs and dwarf galaxies (e.g. \citealt{Janssens2024}) and in line with other works \citep[e.g.][]{Amorisco2018, Saifollahi2022, Saifollahi2025}.

In addition to the spatial distribution of GC candidates, we must also consider their apparent asymmetry (see also \citealt{Marleau2024}). While there is a large number of candidates in the north-west of the galaxy (upper-right of the image, Fig.~\ref{fig:imageDistribution}), the light distribution of the galaxy is symmetrical (Fig.~\ref{fig:rgbgrey} and Sec.~\ref{sec:sbprofs}). This issue is addressed in Appendix \ref{app:asymmetry}, concluding that we cannot rule out that the intriguing distribution of GCs in this galaxy is mere coincidence. Nonetheless, this highlights the importance of analysing the imaging before performing spatial completeness corrections, which implicitly assume symmetry for the GC system. As \dg{} shows, these assumptions might not stand and lead to the incorrect estimation of the number of GCs.

\subsection{The distance to \dg{}}\label{sec:Distance}

\begin{figure*}[]
    \centering
 \includegraphics[width=0.7\textwidth]{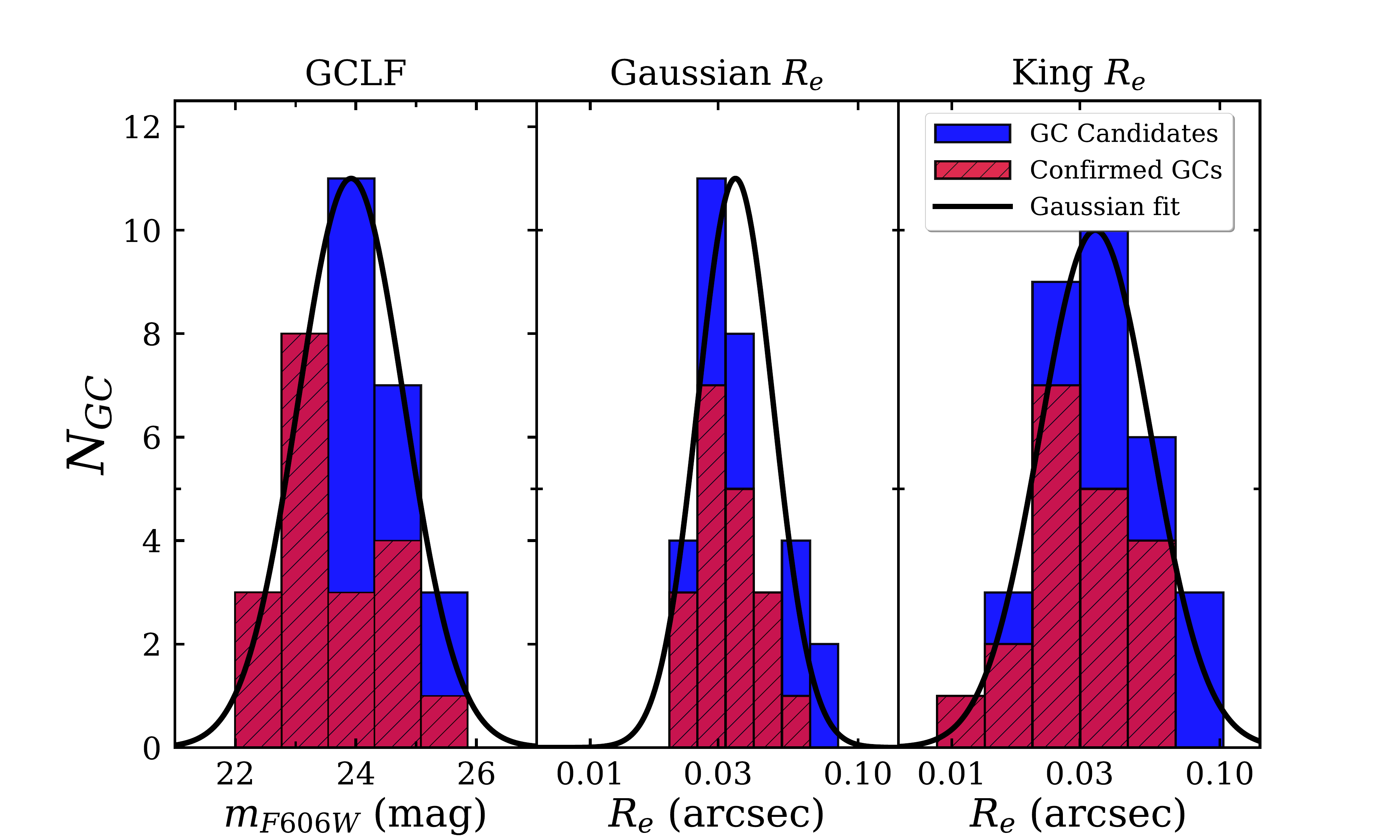}
    \caption{Left panel: Globular Cluster Luminosity Function of \dg. Middle panel: Distribution of $R_e$ assuming that the intrinsic shape of the GC is a Gaussian. Right panel: Distribution of $R_e$ assuming that the intrinsic shape of the GC is a King model. The spectroscopically confirmed GCs from \citet{Muller2020} and \citet{Haacke2025} are indicated in red (hatched), the population identified in this study is in blue. The widths of the bins have been calculated using the Freedman–Diaconis rule \citep{Freedman1981}. The black solid line shows a Gaussian fit to the data.}
    \label{fig:AppendixGCLF}
\end{figure*}

In addition to disagreeing on the number of GCs, different studies also assume different distances to \dg.
The UDG is located in the field-of-view (FOV) of the NGC5846 galaxy group, at a projected distance of $\sim21$\arcmin{} from NGC5846. Given the apparent consistency of the galaxy's velocity with the group \citep{Muller2020, Haacke2025}, this strongly suggests that the distance to \dg{} is the same as to NGC5846 (i.e. $26.5 \pm 0.8$ Mpc from the weighted average distance of the members from Cosmicflows-3 catalogue \citealt{Tully2016, Kourkchi2017, Danieli2022}). However, \citet{Muller2021}, using the GC luminosity function of the GCs (GCLF), finds that \dg{} is closer, located at $20.7^{+2.3}_{-2.1}$ Mpc.

The peak of the GCLF is generally assumed to be universal and, therefore, it has been used extensively to determine distances to galaxies \citep{Rejkuba2012}. Similarly, the $R_e$ distribution of local GCs follow an ---apparently--- universal Gaussian distribution, with no correlation between magnitude and $R_e$. Thus, similar to \cite{Trujillo2019}, using the GCLF and the distribution of $R_e$ we can obtain two independent distance estimates to \dg{}.

In order to use the $R_e$ of the GCs as a distance indicator, we need to accurately estimate the intrinsic $R_{e}$ of the GCs, as it is affected (broadened) by the PSF of the images. Although we already estimated their intrinsic $R_e$ based on the FWHM measurement from \texttt{SExtractor} (See Sec.~\ref{sec:GCSelection}), we improve these measurements to provide a robust estimate of the distance. To do this, we explored both a Gaussian and a King \citep{king1962} model as the intrinsic shapes of a GC, and we use both values below. Further details on the measurement of the $R_{e}$ are given in Appendix \ref{app:re}.

The GCLF and the distribution of $R_{e}$ of the GC candidates are shown in Fig.~\ref{fig:AppendixGCLF}. The candidates that appear to be point-like sources (see Sec.~\ref{sec:Contamination}) have not been included because they are likely interlopers and might bias our distance estimation, thus remaining a total of 32 GCs candidates. The hatched red histogram corresponds to the spectroscopically confirmed GCs from  \citet{Muller2020} and \citet{Haacke2025}, while the blue histogram corresponds to the GC candidates in this study. The black line indicates a Gaussian fit to the combined distributions. For the GCLF, we find that the peak is at $m_{F606W}= 23.92$ mag with a $\sigma = 0.88$. Regarding the $R_e$ distribution, the peak is located at $0\farcs0348$ when fitting a Gaussian model and $0\farcs0343$ for the King model, with standard deviations of $0\farcs0115$ and $0\farcs0170$, respectively.  It is noteworthy that the peaks of the distributions are the same within errors, showing that using these values to infer the distance to the galaxy is robust regardless of the assumed intrinsic profile. The slightly higher dispersion of the King model is likely due to the higher number of free parameters to be fit. The $R_e$ values used for the distributions and for measuring the distance have been circularised\footnote{$R_{e,circ} = R_{e,SMA} \times \sqrt{1 - ellipticity}$, where $R_{e,SMA}$ is the effective radius measured along the semi-major axis}.

In the literature, there are different estimations of the magnitude of the GCLF peak and the $R_{e}$ peak, depending on different factors such as the galaxy type and the metallicities of the GCs. We calculate the distance to \dg{} using different estimates for both the GCLF and $R_{e}$ peaks of the distribution, which are listed in the Appendix \ref{app:dist_calibrations}. Fig.~\ref{Fig:DistancePlot} presents the different distance estimates that are obtained for the different GCLF and $R_{e}$ calibrations explored. The left panel shows the distances assuming a Gaussian model as the intrinsic GC profile, while the right panel assumes a King model. For comparison, the distance obtained by \cite{Muller2021} (brown) and that obtained (grey) and assumed (pink) by \cite{Danieli2022} are shown in a black box at the bottom.
We estimated a weighted mean distance of \gaussianDist{} Mpc, assuming a Gaussian model for the GCs and \kingDist{} Mpc assuming a King model, resulting in an average distance to \dg{} of \meanDist{} Mpc. This distance implies a stellar mass for \dg{} of \stellarMass.

The distance derived in this work is in agreement with the $20.7\pm^{2.3}_{2.1}$ Mpc obtained in \cite{Muller2021} using the GCLF and with the $21\pm5$ Mpc by \cite{Danieli2022} using surface brightness fluctuations. 
However, our estimated distance is not in agreement with the distance to the NGC 5846 group of galaxies ($26.5 \pm 0.8$). Nonetheless, the galaxies in the group have estimated distances ranging from 22.7 Mpc (NGC 5839) to 34.5 Mpc (NGC 5838) \citep{Kourkchi2017}. Therefore, \dg{} could be marginally associated to the group even though the estimated distance puts the galaxy well beyond its virial radius \citep[0.6 Mpc, ][]{Kourkchi2017}.

\begin{figure*}
    \centering\includegraphics[width=0.7\linewidth]{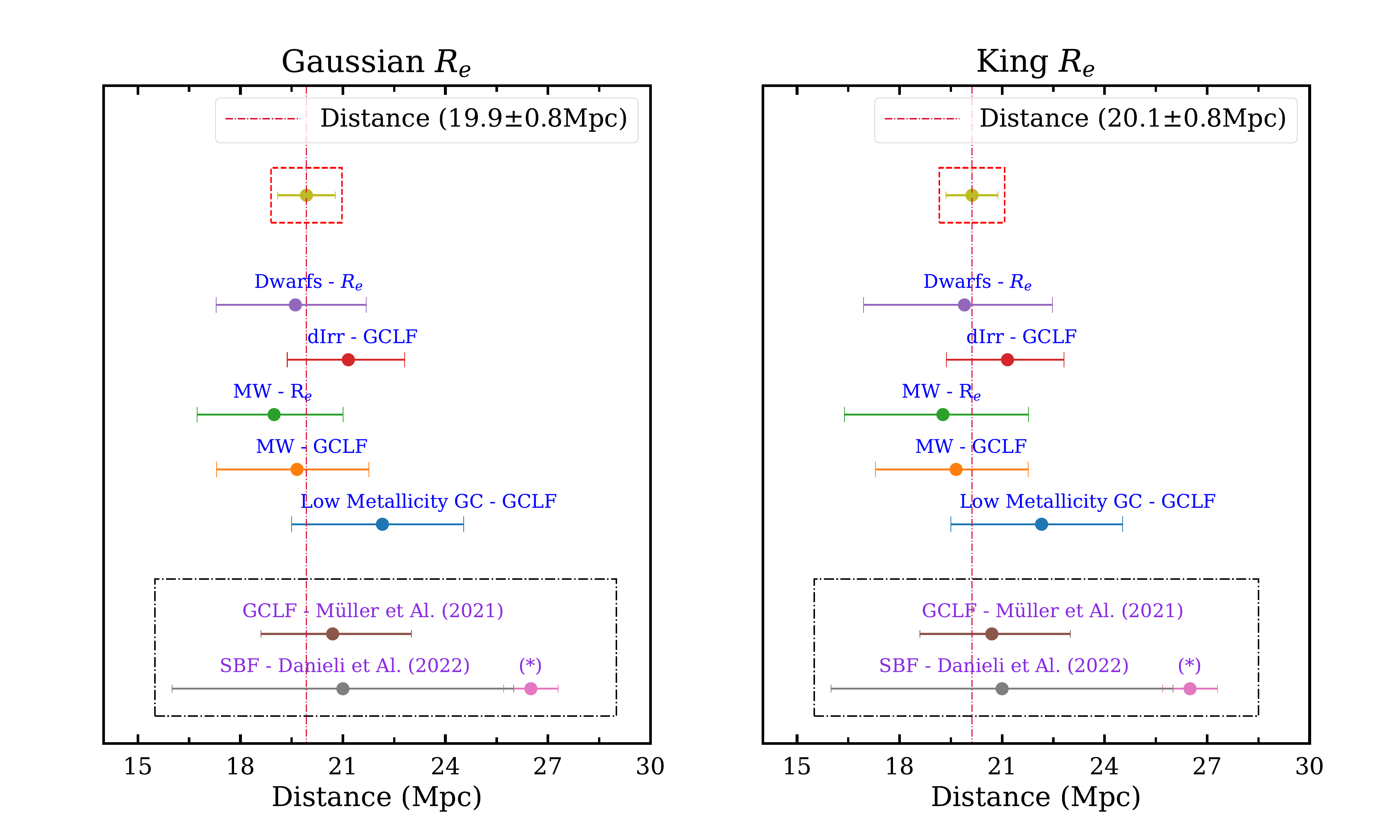}
    \caption{Distance to \dg. The left panel shows the different estimates where the $R_{e,GC}$ is obtained assuming that the GC intrinsic profile follows a  Gaussian model, while the right panel assumes a King model. Each of the points shows a different method or calibration used to estimate the distance to \dg{} as labelled.    
    The grey box separates the distances obtained in \cite{Muller2021} and in \cite{Danieli2022}. The (*) indicates the distance assumed in \cite{Danieli2022} based on the distance to the NGC5846 group. In the red box, we plot the weighted mean distance  obtained using the different estimates for the GCLF and $R_{e,GC}$ methods.}
    \label{Fig:DistancePlot}
\end{figure*}

\subsection{The halo mass of \dg{}}

The number of GCs of a galaxy is closely linked to its halo mass, providing us a way to estimate the mass of its dark matter halo. This GC-Halo mass relation has been proposed to hold for a wide range of halo masses, from $\sim 10^9 M_\odot$ to $\sim 10^{14} M_\odot$ \citep{Forbes2018}. Different authors have calibrated this GC-Halo mass relationship. For example, \citet{Harris2017} determined a constant ratio between the mass in the GC system and the total halo mass of $M_{GCs}/M_h = 2.9\times10^{-5}$. If we assume a mean GC mass of $1.0\times10^5 M_\odot$ \citep{Harris2017}, we obtain a total mass in the GC system of (\finalPopulation)$\times10^{5} M_\odot$, and a halo mass of \totalMass. This estimation is consistent with \citet{Muller2021} and \citet{Marleau2024} ($(0.9 \pm 0.2)\times10^{11} M_\odot$ and $(1.29^{+0.26}_{-0.22})\times10^{11} M_\odot$ respectively). The number of GCs and halo mass derived here for \dg{} are in line with those of other UDGs \citep[e.g.][]{BeasleyTrujillo2016, Beasley2016, Amorisco2018, Saifollahi2022, Mihos_2025}.

Consequently, \dg{} does not seem to inhabit an unexpectedly massive halo, in line with the halo masses expected for a dwarf galaxy (up to $\sim 10^{10-11} M_\odot$). However, for the halo mass we derive here, we would expect a stellar mass of $\sim 10^{9} M_\odot$ \citep[e.g.][]{Girelli2020} rather than what we find (\stellarMass). This lower-than-expected stellar masses are observed in other UDGs, indicating rapid star formation quenching after the formation of the GC systems of these galaxies, as suggested in \citet{BeasleyTrujillo2016}.

\section{Conclusions}

\dg{} is an UDG with an unusually large GC population. There is an ongoing debate about how large this population is, where previous estimations of its GC system range from less than 30 up to more than 50 GCs. In this work, we use the multi-wavelength capabilities and high spatial resolution of archival HST data, and deep GTC data to provide a clean sample of GC candidates of this galaxy. Additionally, we explore the stellar body of the galaxy, comparing it with the distribution of GCs. Finally, we do a careful analysis of the distance to the galaxy.

We find that the GC population of \dg{} is \finalPopulation{} GCs, which implies a halo mass of \totalMass. This result is in agreement with the previous lower estimates of the GC system of the galaxy (\citealt{Muller2021, Marleau2024}). Nevertheless, these values are larger than expected for a galaxy of the stellar mass of \dg{} ($M_*=9.2^{+3.0}_{-2.3} \times 10^7$ M$_\odot$ at a distance of $20.7^{+2.3}_{-2.1}$ Mpc or, alternatively, $M_*=1.5 \pm 0.2 \times 10^8$ M$_\odot$ at $26.5\pm0.8$ Mpc). This discrepancy suggests that the galaxy may have undergone rapid quenching of the stellar body of the galaxy after the bulk of GC formation.

We also confirm the previously reported asymmetry in the distribution of GCs, with a clear over-density towards the north-western region of the galaxy. Interestingly, there is no trace of asymmetry in the body of the host galaxy. After performing a statistical analysis, we conclude that this GC asymmetry is not significant, being it compatible with a random distribution of the GC system. Additionally, the GC system appears to be highly concentrated on top of the galaxy, having $80\%$ of the GCs inside one effective radius. Thus, the stellar light component of the galaxy is more extended that its GC system, having a $R_{e,GC}/R_e$ of 0.7. Finally, we also estimate a distance of \meanDist{} Mpc to \dg{} based on the peak of the GCLF and peak of the distribution of the effective radius of the GCs.

This work emphasises the necessity of using multi-wavelength and high spatial resolution data to minimise foreground and background contamination in the GC systems of UDGs. We also warn against assuming symmetry in the distribution of GCs when making spatial or coverage corrections to infer the number of GCs of these low mass systems.

\begin{acknowledgements}
We thank the referee for the constructive comments that helped improve the quality of the manuscript. We thank F. Marleau for providing the GC data of MATLAS2019 from \citet{Marleau2024}. Also thank Lydia Haacke and Duncan A. Forbes for letting us compare with their data prior to publication \citep{Haacke2025}.
SGA acknowledges support from grant PID2022-140869NB-I00 from the Spanish Ministry of Science and Innovation. MM acknowledges support from the project PCI2021-122072-2B, financed by MICIN/AEI/10.13039/501100011033, and the European Union “NextGenerationEU”/RTRP, from grant RYC2022-036949-I financed by the MICIU/AEI/10.13039/501100011033 and by ESF+ and program Unidad de Excelencia Mar\'{i}a de Maeztu CEX2020-001058-M. G.G. acknowledges support from IAC project P/302304 and through the PID2022-140869NB-I00 grant from the Spanish Ministry of Science and Innovation which is partially supported through the state budget and the regional budget of the Consejería de Economía, Industria, Comercio y Conocimientoof the Canary Islands Autonomous Community. IT acknowledges support from the State Research Agency (AEI-MCINN) of the Spanish Ministry of Science and Innovation under the grant PID2022-140869NB-I00, financed by the Ministry of Science and Innovation, through the State Budget and by the Canary Islands Department of Economy, Knowledge, and Employment, through the Regional Budget of the Autonomous Community. This research also acknowledge support from the European Union through the following grants: "UNDARK" and "Excellence in Galaxies - Twinning the IAC" of the EU Horizon Europe Widening Actions  programmes (project numbers 101159929 and 101158446). Funding for this work/research was provided by the European Union (MSCA EDUCADO, GA 101119830). Views and opinions expressed are however those of the author(s) only and do not necessarily reflect those of the European Union or European Research Executive Agency (REA). MM, GG and IT acknowledge support from IAC project P/302302.

This work makes use of the following code:
\texttt{astropy} \citep{Astropy},  
          \texttt{Source Extractor} \citep{BertinArnouts1996},
          \texttt{Gnuastro}  \citep{Akhlaghi2015},     
          \texttt{photutils} \citep{larrybradley2023},
          \texttt{numpy} \citep{harris2020array},
          \texttt{scipy} \citep{2020SciPy-NMeth},
          \texttt{Imfit} \citep{imfit2015}.
\end{acknowledgements}

\bibliography{sergio.bib}

\begin{appendix}

\section{Effect on the results by including the anomalous globular cluster}\label{app:anomalous}
\begin{figure}[H]
    \centering
    \includegraphics[width=0.9\columnwidth]{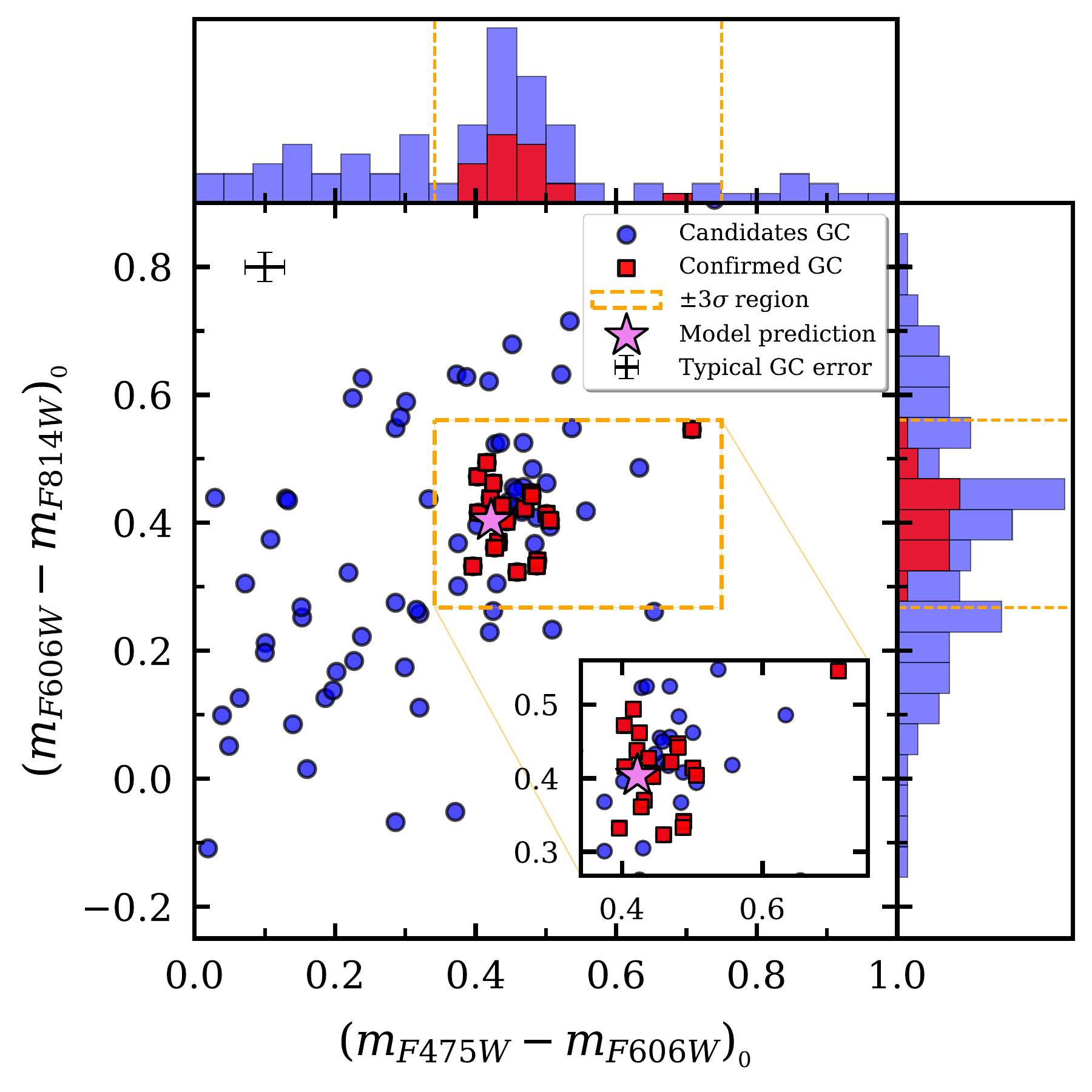}
    \caption{Same as Fig. \ref{fig:colourColourDiagram} but extending the selection region in the $(m_{F475W} - m_{F606W})_0$ colour up to  0.75 for including the anomalous GC.}
    \label{fig:colourColourDiagram_withAnomalous}
\end{figure}

In Sec. \ref{sec:colour-colour}, we find that one of the GC confirmed by \citet{Haacke2025} shows colours that are very different to the rest of confirmed GCs, and we did not consider it as bona fide GC in the subsequent analysis. In this section, we test the robustness of our analysis by including it.  Fig. \ref{fig:colourColourDiagram_withAnomalous} shows the resulting colour-colour diagram. We widened the selection box in the $(m_{F475W} - m_{F606W})_0$ colour up to 0.75, thus including the 20 spectroscopically confirmed objects from \cite{Haacke2025}. With this selection region, the number of GC candidates results in 41, 3 more than in Sec. \ref{sec:colour-colour}, including the spectroscopically confirmed.

\begin{figure}
    \centering
\includegraphics[width=1.05\linewidth]{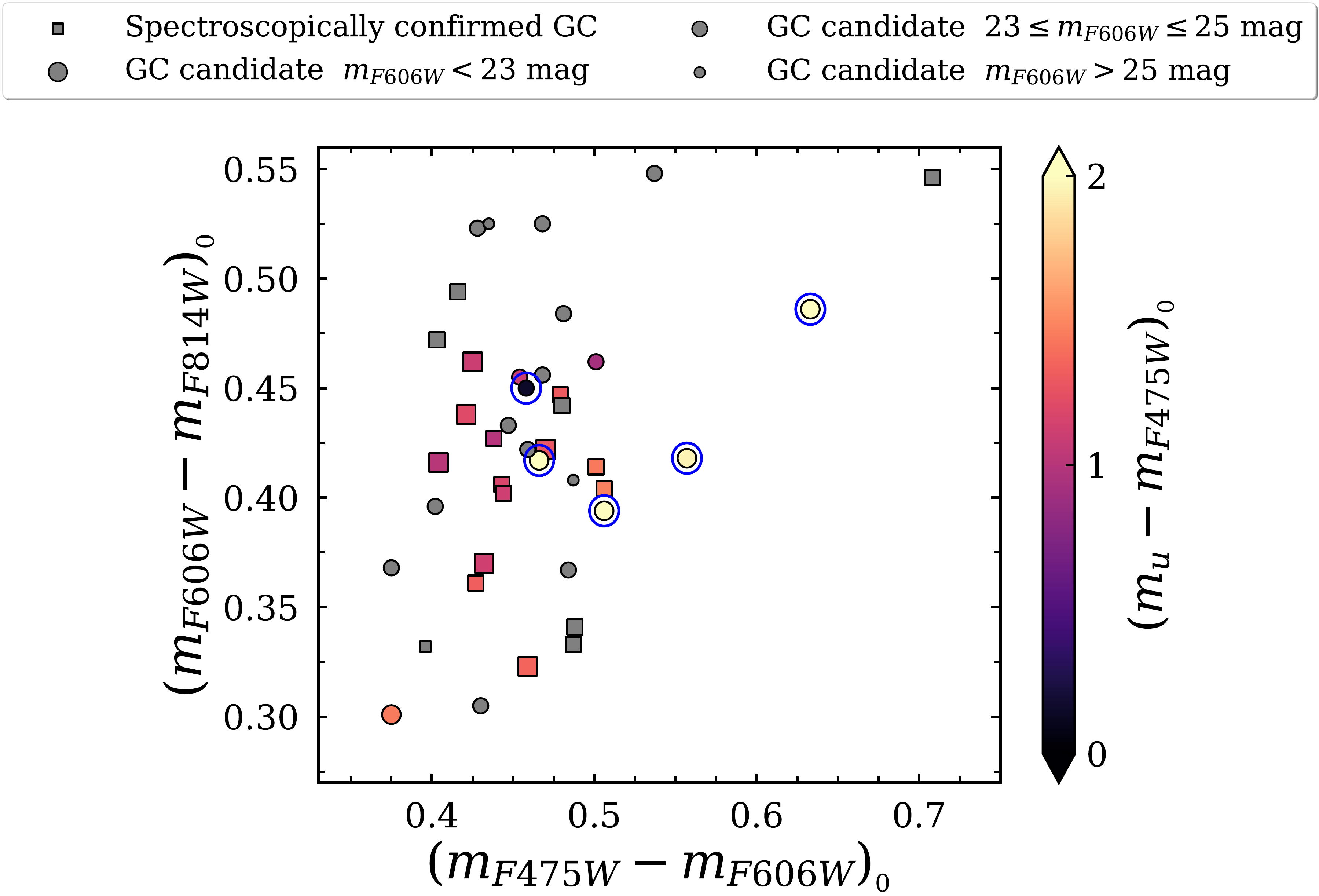}
    \caption{Same as Fig. \ref{fig:colourDiagram_u} but applying a wider selection region to include the anomalous GC. Most of the new candidates have very red colours and are rejected due to their $(m_u-m_{F475W})_0$ values.}
    \label{fig:colourDiagram_u_anomalous}
\end{figure}

\begin{figure}
    \centering
\includegraphics[width=0.7\linewidth]{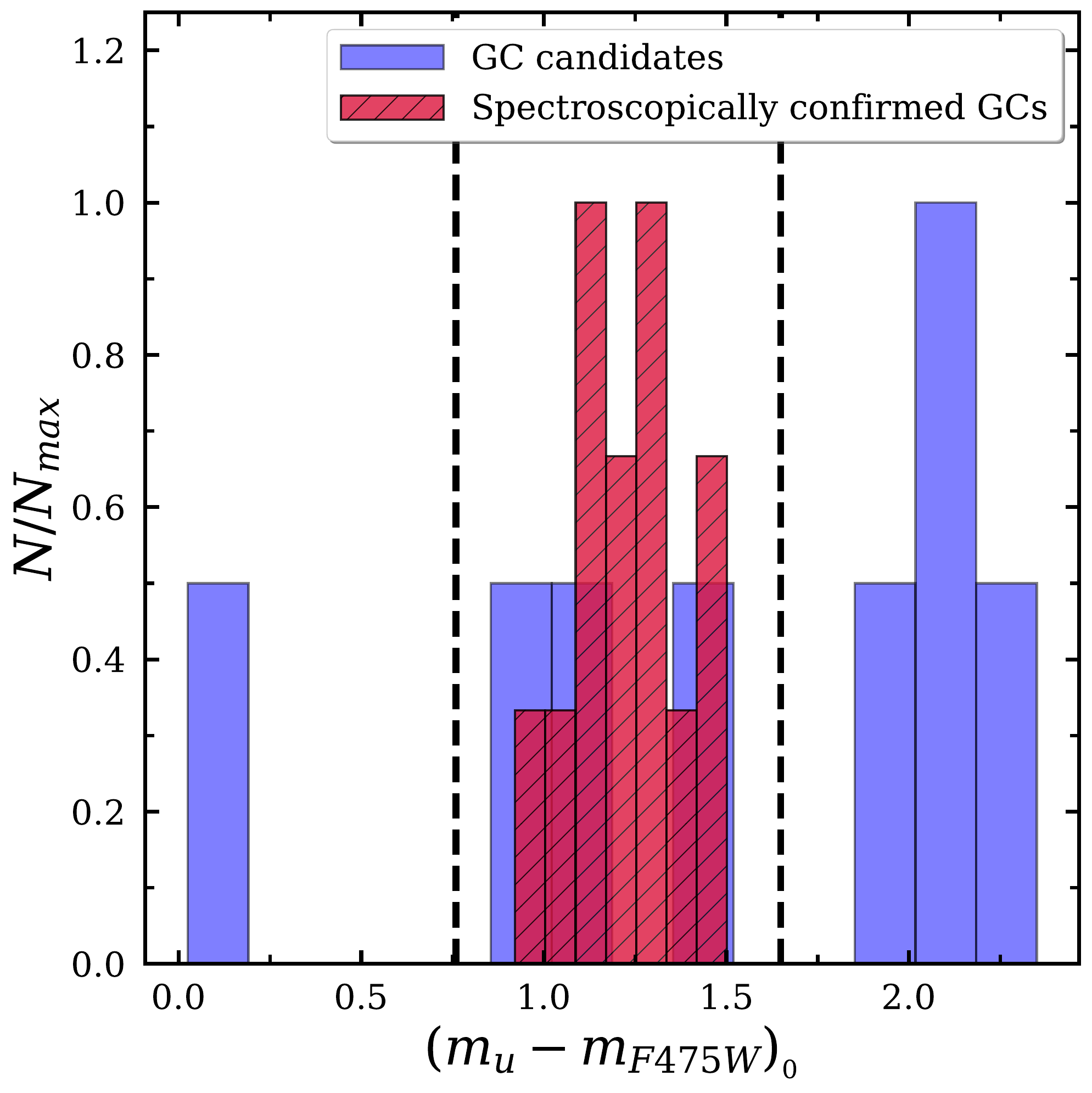}
    \caption{Same as Fig. \ref{fig:u-475Hist} but applying a wider selection region in the $(m_{F475W} - m_{F606W})_0$ colour for including the anomalous GC.}
    \label{fig:u475Hist_anomalous}
\end{figure}

Following the same methodology, we also use the $(m_u-m_{F475W})_0$ colour to further clean GC candidates. Fig. \ref{fig:colourDiagram_u_anomalous} shows the colour-colour diagram with the three colours (similar to Fig. \ref{fig:colourDiagram_u}) and Fig. \ref{fig:u475Hist_anomalous} shows the distribution of the $(m_u-m_{F475W})_0$ colours (similar to Fig. \ref{fig:u-475Hist}). None of the additional GC candidates has a $(m_u-m_{F475W})_0$ colour compatible with our criterion. The anomalous spectroscopically confirmed object is too faint in the $u$-band for reliably getting its $(m_u-m_{F475W})_0$ colour. Therefore, at this point we have 36 GC candidates, only the spectroscopically confirmed object is added to the sample.

Following the same steps in this work, we statistically characterise the GC population, i.e. perturbing the measured parameters of the candidates based on the errors obtained from the completeness tests (see. Sec. \ref{sec:finalnumber} and Appendix \ref{app:completeness}). The expected population of GCs results in $35 \pm 3$. After adding the completeness correction due to the limited depth of the data, we end up with $36\pm3$. To summarise, if we consider all the 20 spectroscopically confirmed GCs, we add one extra spectroscopically confirmed GC, and 3 extra objects in the expected population of GCs (from $32\pm3$ to $35\pm3$ GCs candidates). 

Remarkably, the final number of objects does not significantly change even if the selection range in the colour $(m_{F475W} - m_{F606W})_0$ is twice as large as in the main analysis. As stated in Sec. \ref{sec:comparisonWithStudies} when comparing with the results from \cite{Marleau2024} and especially from \cite{Danieli2022}, the colour range is really important for an adequate selection of GC candidates. However, the confirmed GC is located in a relatively clean region in colour space (Fig. \ref{fig:colourColourDiagram_withAnomalous}), making our results robust whether we include this object or not.

The measured properties of the confirmed anomalous object are listed in the last row of Table \ref{tab:gcPopulation}. It is remarkable that, even though the profile and the fit of a Gaussian/King profile of this object look fairly normal, it shows large values of $\chi^{2}_{red}$. Neither the imaging nor the results of the fits provide an obvious explanation to its anomaly.

\section{Photometric and morphological accuracy}\label{app:completeness}

\begin{figure*}
    \centering
    \includegraphics[width=0.8\linewidth, trim=40 40 40 40, clip]{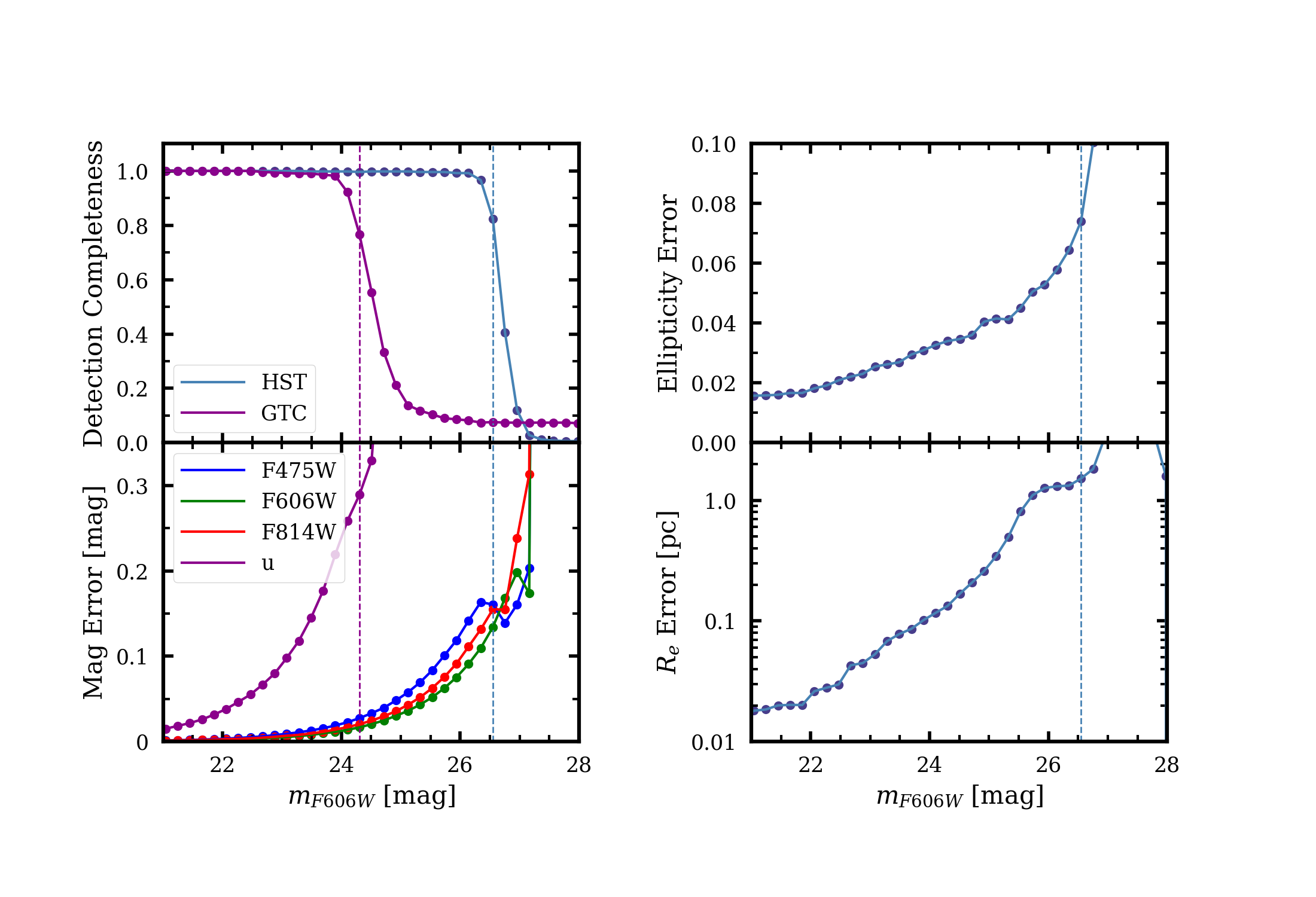}
    \caption{Results of the tests of the injection of mock GCs. Top left: Completeness of the HST data when simultaneously detecting the injected GCs in the three HST bands (blue line) and for the OSIRIS+ $u$-band (purple line). Bottom left: photometric errors for all the bands used in characterizing the GCs of \dg. Top right: ellipticity errors in the WFC3/\textit{F606W}, where the ellipticity is measured. Bottom right: $R_{e}$ errors in the WFC3/\textit{F606W}, converted to pc assuming a distance of 20.7 Mpc. The dashed vertical lines indicates when the detections fall below $90\%$.}
    \label{fig:simulations} 
\end{figure*}

To assess the completeness, photometric and morphological errors of our GC sample, we have performed tests in which we inject, retrieve and analyse artificial GC-like sources. These artificial GC-like sources have the colours of the spectroscopically confirmed GCs (i.e. $(m_{F475W} - m_{F606W})_0$ = 0.44, $(m_{F606W} - m_{F814W})_0$ = 0.41, and $(m_u-m_{F475W})_0$ = 1.20, see Sec.~\ref{sec:colour-colour} for details). The mock GCs were subjected to the same analysis as the real data. 
For the HST data, we model the GCs by circularising the mean profile of the spectroscopically confirmed GCs (therefore having an ellipticity = 0 and FWHM = 2.7 px, see Sec.~\ref{sec:GCprofiles}). For the $u$-band data, the model is based on a PSF constructed by circularising the radial profile of bright and non-saturated stars in the image (ellipticity = 0 and FWHM = 4.3 px, see Sec. \ref{u-band_phot}). The profiles have been extended with the same power-law used for correcting the photometry in the analysis of real sources. We injected these artificial GCs (500 per magnitude bin) randomly distributed throughout the image with magnitudes between 21 and 28. Fig.~\ref{fig:simulations} show the results of these tests. The detection completeness represents the rate of recovered simulated GCs per magnitude bin (upper left panel). We have also compared the injected and recovered values of magnitude (lower left), ellipticity (upper right) and $R_e$ (lower right; given in pc assuming a distance of 20.7 Mpc).

In the HST data, the completeness limit (for objects simultaneously detected in the three filters) drops below 90\% at $m_\textit{F606W} \sim$ 26.5 mag (blue dashed line in Fig. \ref{fig:simulations}). The errors at this magnitude are : $\sim0.15$ mag in photometry, $0.08$ in ellipticity and 1.5 pc in $R_e$.
In the case of the $u$-band, the completeness drops below 90\% at $m_u \sim$ 25.8 mag (corresponding to $m_\textit{F606W} \sim$ 24.3 mag, purple dashed line). The photometric error in the $u$-band at this magnitude is 0.3 mag.

\section{Properties of the GC candidates of \dg}

Table \ref{tab:gcPopulation} lists the properties of the confirmed and candidates GCs of \dg. Magnitudes are corrected from Galactic extinction and aperture corrections as described in Sec.~\ref{sec:GCSelection}. Candidates with no given magnitude in the $u$-band are too faint to be measured in the $u$-band. The magnitude errors are obtained from the completeness tests (Appendix \ref{app:completeness}) and combined with the uncertainty in the correction due to finite apertures.

The errors for the effective radius based on \sextractor{} are based on the mock GC tests (Appendix \ref{app:completeness}) and the error for the values obtained from fitting Gaussian and King profiles are obtained from the typical difference between the Gaussian $R_e$ and the King $R_e$.

\begin{table*}[htbp]
    \centering
    \caption{Properties of the GC candidates identified in \dg. Magnitudes are corrected from Galactic extinction and aperture corrections (see Sec. \ref{sec:AperAndCorrect}, \ref{sec:GCSelection}, and \ref{u-band_phot} for details). The $R_{e,Gauss,SEx}$ column indicates the $R_e$ measurement obtained from SExtractor as described in Sec. \ref{sec:GCSelection}. The following four columns are the measurements obtained by fitting the GCs with an intrinsic Gaussian profile ($ R_{e,Gauss}$, $\chi^{2}_{red, Gauss}$) and an intrinsic King profile ($R_{e,King}$  $\chi^{2}_{red, King}$) as described in Sec. \ref{app:re}.}
    \label{tab:gcPopulation}
    \begin{adjustbox}{width=1\textwidth}
    \begin{tabular}{cccccccccccccc}
    \toprule \toprule
	Name & RA & Dec & F475W & F606W & F814W & u & $R_{e,Gauss,SEx}$ & $ R_{e,Gauss}$ & $\chi^{2}_{red, Gauss}$ & $R_{e,King}$ & $\chi^{2}_{red, King}$ & Ellipticity \\
	 & (J2000.0) & (J2000.0) & (mag) & (mag) & (mag) & (mag) & (arcsec) & (arcsec) & & (arcsec) & &   \\
	\midrule
	$\,\,\,$GC1$^*$	& 226.334535  &  1.812961  & 22.49  $\pm$  0.02  & 22.09  $\pm$  0.01  & 21.67  $\pm$  0.01  & 23.50  $\pm$  0.03  & 0.056  $\pm$  0.001  & $0.045  \pm  $0.006  & 3.20  & $0.063  \pm  $0.006  & 1.34  & 0.10  $\pm$  0.02\\[0.15cm]
	$\,\,\,$GC2$^*$	& 226.333920  &  1.812404  & 22.82  $\pm$  0.02  & 22.40  $\pm$  0.01  & 21.96  $\pm$  0.01  & 24.02  $\pm$  0.05  & 0.047  $\pm$  0.001  & $0.035  \pm  $0.006  & 1.86  & $0.034  \pm  $0.006  & 1.23  & 0.10  $\pm$  0.02\\[0.15cm]
	$\,\,\,$GC3$^*$	& 226.336544  &  1.817503  & 23.15  $\pm$  0.02  & 22.69  $\pm$  0.01  & 22.36  $\pm$  0.02  & 24.50  $\pm$  0.08  & 0.040  $\pm$  0.002  & $0.028  \pm  $0.006  & 1.42  & $0.021  \pm  $0.006  & 1.32  & 0.09  $\pm$  0.02\\[0.15cm]
	$\,\,\,$GC4$^*$	& 226.335655  &  1.811625  & 23.26  $\pm$  0.03  & 22.84  $\pm$  0.02  & 22.38  $\pm$  0.02  & 24.36  $\pm$  0.07  & 0.047  $\pm$  0.002  & $0.038  \pm  $0.006  & 1.55  & $0.050  \pm  $0.006  & 1.08  & 0.07  $\pm$  0.02\\[0.15cm]
	$\,\,\,$GC5$^*$	& 226.335160  &  1.813678  & 23.39  $\pm$  0.03  & 22.92  $\pm$  0.02  & 22.49  $\pm$  0.02  & 24.66  $\pm$  0.09  & 0.042  $\pm$  0.002  & $0.032  \pm  $0.006  & 1.26  & $0.029  \pm  $0.006  & 1.08  & 0.07  $\pm$  0.02\\[0.15cm]
	$\,\,\,$GC6$^*$	& 226.333500  &  1.811074  & 23.41  $\pm$  0.03  & 22.98  $\pm$  0.02  & 22.61  $\pm$  0.02  & 24.53  $\pm$  0.08  & 0.040  $\pm$  0.002  & $0.028  \pm  $0.006  & 1.20  & $0.024  \pm  $0.006  & 1.10  & 0.06  $\pm$  0.02\\[0.15cm]
	$\,\,\,$GC7$^*$	& 226.333823  &  1.810629  & 23.46  $\pm$  0.03  & 23.02  $\pm$  0.02  & 22.62  $\pm$  0.02  & 24.64  $\pm$  0.09  & 0.040  $\pm$  0.002  & $0.031  \pm  $0.006  & 1.67  & $0.028  \pm  $0.006  & 1.58  & 0.05  $\pm$  0.02\\[0.15cm]
	$\,\,\,$GC8$^{**}$	& 226.331357  &  1.815125  & 23.62  $\pm$  0.03  & 23.18  $\pm$  0.02  & 22.78  $\pm$  0.02  & 24.74  $\pm$  0.10  & 0.043  $\pm$  0.002  & $0.034  \pm  $0.006  & 1.40  & $0.037  \pm  $0.006  & 1.22  & 0.07  $\pm$  0.03\\[0.15cm]
	$\,\,\,$GC9$^*$	& 226.336456  &  1.815258  & 23.61  $\pm$  0.03  & 23.13  $\pm$  0.02  & 22.69  $\pm$  0.02  & 24.92  $\pm$  0.12  & 0.036  $\pm$  0.002  & $0.024  \pm  $0.006  & 1.11  & $0.020  \pm  $0.006  & 1.07  & 0.06  $\pm$  0.03\\[0.15cm]
	$\,\,\,$GC10$^*$	& 226.329937  &  1.811504  & 24.02  $\pm$  0.03  & 23.52  $\pm$  0.02  & 23.10  $\pm$  0.02  & 25.46  $\pm$  0.21  & 0.043  $\pm$  0.003  & $0.037  \pm  $0.006  & 1.20  & $0.039  \pm  $0.006  & 1.09  & 0.07  $\pm$  0.03\\[0.15cm]
	$\,\,\,$GC11$^*$	& 226.334080  &  1.810156  & 24.06  $\pm$  0.03  & 23.63  $\pm$  0.02  & 23.27  $\pm$  0.02  & 25.38  $\pm$  0.19  & 0.045  $\pm$  0.003  & $0.030  \pm  $0.006  & 1.09  & $0.024  \pm  $0.006  & 1.06  & 0.05  $\pm$  0.03\\[0.15cm]
	$\,\,\,$GC12$^*$	& 226.335797  &  1.813603  & 24.00  $\pm$  0.03  & 23.50  $\pm$  0.02  & 23.09  $\pm$  0.02  & 25.48  $\pm$  0.21  & 0.039  $\pm$  0.003  & $0.024  \pm  $0.006  & 1.12  & $0.018  \pm  $0.006  & 1.11  & 0.09  $\pm$  0.03\\[0.15cm]
	$\,\,\,$GC13$^*$	& 226.335548  &  1.812565  & 24.00  $\pm$  0.03  & 23.60  $\pm$  0.02  & 23.12  $\pm$  0.02  & ... & 0.036  $\pm$  0.003  & $0.022  \pm  $0.006  & 1.23  & $0.013  \pm  $0.006  & 1.23  & 0.04  $\pm$  0.03\\[0.15cm]
	$\,\,\,$GC14$^*$	& 226.337543  &  1.817332  & 24.81  $\pm$  0.05  & 24.33  $\pm$  0.03  & 23.88  $\pm$  0.03  & ... & 0.056  $\pm$  0.005  & $0.044  \pm  $0.006  & 1.08  & $0.046  \pm  $0.006  & 1.06  & 0.08  $\pm$  0.03\\[0.15cm]
	$\,\,\,$GC15$^*$	& 226.335041  &  1.810901  & 24.80  $\pm$  0.05  & 24.36  $\pm$  0.03  & 23.93  $\pm$  0.03  & ... & 0.044  $\pm$  0.006  & $0.031  \pm  $0.006  & 1.11  & $0.034  \pm  $0.006  & 1.08  & 0.11  $\pm$  0.03\\[0.15cm]
	$\,\,\,$GC16$^*$	& 226.332882  &  1.812673  & 24.71  $\pm$  0.05  & 24.22  $\pm$  0.03  & 23.88  $\pm$  0.03  & ... & 0.043  $\pm$  0.005  & $0.031  \pm  $0.006  & 1.07  & $0.028  \pm  $0.006  & 1.06  & 0.09  $\pm$  0.03\\[0.15cm]
	$\,\,\,$GC17$^*$	& 226.338345  &  1.818353  & 25.11  $\pm$  0.06  & 24.70  $\pm$  0.04  & 24.20  $\pm$  0.04  & ... & 0.065  $\pm$  0.008  & $0.046  \pm  $0.006  & 1.02  & $0.062  \pm  $0.006  & 1.00  & 0.06  $\pm$  0.04\\[0.15cm]
	$\,\,\,$GC18$^*$	& 226.333652  &  1.815722  & 25.18  $\pm$  0.06  & 24.69  $\pm$  0.04  & 24.36  $\pm$  0.04  & ... & 0.046  $\pm$  0.008  & $0.035  \pm  $0.006  & 1.02  & $0.031  \pm  $0.006  & 1.01  & 0.05  $\pm$  0.04\\[0.15cm]
	$\,\,\,$GC19$^*$	& 226.334855  &  1.810012  & 25.98  $\pm$  0.10  & 25.58  $\pm$  0.07  & 25.25  $\pm$  0.08  & ... & 0.084  $\pm$  0.035  & $0.058  \pm  $0.006  & 1.14  & $0.062  \pm  $0.006  & 1.14  & 0.16  $\pm$  0.05\\[0.15cm]
	$\,\,\,$GC20$^{a, b}$	& 226.349708  &  1.827662  & 24.85  $\pm$  0.05  & 24.48  $\pm$  0.03  & 24.11  $\pm$  0.04  & ... & 0.020  $\pm$  0.007  & $0.011  \pm  $0.006  & 1.22  & $0.013  \pm  $0.006  & 1.23  & 0.05  $\pm$  0.03\\[0.15cm]
	$\,\,\,$GC21$^{a}$	& 226.335338  &  1.821988  & 24.64  $\pm$  0.04  & 24.18  $\pm$  0.03  & 23.76  $\pm$  0.03  & ... & 0.047  $\pm$  0.005  & $0.022  \pm  $0.006  & 1.09  & $0.017  \pm  $0.006  & 1.09  & 0.12  $\pm$  0.03\\[0.15cm]
	GC22	& 226.330051  &  1.814242  & 24.44  $\pm$  0.04  & 23.95  $\pm$  0.02  & 23.47  $\pm$  0.02  & ... & 0.038  $\pm$  0.004  & $0.031  \pm  $0.006  & 1.12  & $0.032  \pm  $0.006  & 1.07  & 0.03  $\pm$  0.03\\[0.15cm]
	GC23	& 226.332685  &  1.813938  & 25.34  $\pm$  0.07  & 24.93  $\pm$  0.04  & 24.54  $\pm$  0.05  & ... & 0.047  $\pm$  0.011  & $0.033  \pm  $0.006  & 1.00  & $0.028  \pm  $0.006  & 0.99  & 0.12  $\pm$  0.04\\[0.15cm]
	GC24	& 226.331124  &  1.813442  & 24.64  $\pm$  0.04  & 24.21  $\pm$  0.03  & 23.69  $\pm$  0.03  & ... & 0.046  $\pm$  0.005  & $0.036  \pm  $0.006  & 1.07  & $0.034  \pm  $0.006  & 1.05  & 0.11  $\pm$  0.03\\[0.15cm]
	GC25	& 226.330619  &  1.812414  & 24.77  $\pm$  0.05  & 24.28  $\pm$  0.03  & 23.92  $\pm$  0.03  & ... & 0.042  $\pm$  0.005  & $0.028  \pm  $0.006  & 1.06  & $0.026  \pm  $0.006  & 1.06  & 0.11  $\pm$  0.03\\[0.15cm]
	GC26	& 226.328714  &  1.811063  & 24.95  $\pm$  0.05  & 24.48  $\pm$  0.03  & 23.96  $\pm$  0.03  & ... & 0.078  $\pm$  0.007  & $0.071  \pm  $0.006  & 1.09  & $0.078  \pm  $0.006  & 1.07  & 0.12  $\pm$  0.03\\[0.15cm]
	GC27	& 226.331864  &  1.811113  & 24.48  $\pm$  0.04  & 24.02  $\pm$  0.02  & 23.57  $\pm$  0.03  & 25.58  $\pm$  0.23  & 0.063  $\pm$  0.004  & $0.030  \pm  $0.006  & 1.27  & $0.035  \pm  $0.006  & 1.23  & 0.17  $\pm$  0.03\\[0.15cm]
	GC28	& 226.336051  &  1.812232  & 26.03  $\pm$  0.11  & 25.60  $\pm$  0.07  & 25.07  $\pm$  0.07  & ... & 0.093  $\pm$  0.036  & $0.071  \pm  $0.006  & 2.39  & $0.082  \pm  $0.006  & 2.39  & 0.03  $\pm$  0.05\\[0.15cm]
	GC29	& 226.331104  &  1.810617  & 24.56  $\pm$  0.04  & 24.06  $\pm$  0.02  & 23.60  $\pm$  0.03  & 25.48  $\pm$  0.21  & 0.068  $\pm$  0.005  & $0.060  \pm  $0.006  & 1.13  & $0.069  \pm  $0.006  & 1.09  & 0.12  $\pm$  0.03\\[0.15cm]
	GC30	& 226.330057  &  1.810117  & 24.44  $\pm$  0.04  & 23.99  $\pm$  0.02  & 23.56  $\pm$  0.03  & ... & 0.047  $\pm$  0.004  & $0.039  \pm  $0.006  & 1.12  & $0.038  \pm  $0.006  & 1.09  & 0.14  $\pm$  0.03\\[0.15cm]
	GC31	& 226.332186  &  1.810756  & 26.01  $\pm$  0.11  & 25.52  $\pm$  0.06  & 25.11  $\pm$  0.07  & ... & 0.085  $\pm$  0.032  & $0.060  \pm  $0.006  & 4.18  & $0.059  \pm  $0.006  & 4.18  & 0.16  $\pm$  0.04\\[0.15cm]
	$\,\,\,$GC32$^{a}$	& 226.315201  &  1.804219  & 24.66  $\pm$  0.04  & 24.19  $\pm$  0.03  & 23.74  $\pm$  0.03  & ... & 0.047  $\pm$  0.005  & $0.036  \pm  $0.006  & 1.07  & $0.040  \pm  $0.006  & 1.03  & 0.15  $\pm$  0.03\\[0.15cm]
	$\,\,\,$GC33$^{b}$	& 226.332478  &  1.808550  & 24.89  $\pm$  0.05  & 24.46  $\pm$  0.03  & 24.16  $\pm$  0.04  & ... & 0.025  $\pm$  0.006  & $0.016  \pm  $0.006  & 1.13  & $0.013  \pm  $0.006  & 1.14  & 0.06  $\pm$  0.03\\[0.15cm]
	GC34	& 226.340596  &  1.809073  & 25.27  $\pm$  0.06  & 24.73  $\pm$  0.04  & 24.18  $\pm$  0.04  & ... & 0.080  $\pm$  0.008  & $0.069  \pm  $0.006  & 1.04  & $0.094  \pm  $0.006  & 1.03  & 0.09  $\pm$  0.04\\[0.15cm]
	$\,\,\,$GC35$^{a, b}$	& 226.332256  &  1.802822  & 22.25  $\pm$  0.02  & 21.88  $\pm$  0.01  & 21.58  $\pm$  0.01  & 23.70  $\pm$  0.04  & 0.024  $\pm$  0.001  & $0.007  \pm  $0.006  & 2.38  & $0.013  \pm  $0.006  & 3.18  & 0.06  $\pm$  0.02\\[0.15cm]

    \bottomrule

    \noalign{\vskip 0.3cm}

	$\,\,\,$GC36$^{***}$	& 226.334017  &  1.811001  & 25.68  $\pm$  0.08  & 24.97  $\pm$  0.04  & 24.43  $\pm$  0.04  & ... & 0.078  $\pm$  0.011  & $0.043  \pm  $0.006  & 22.49  & $0.040  \pm  $0.006  & 22.49  & 0.09  $\pm$  0.04\\[0.15cm]

    \end{tabular}
    \end{adjustbox}
\\[0.2cm]
\footnotesize{\textit{$^*$ Spectroscopically confirmed GCs from \citet{Haacke2025}.}}\\
\footnotesize{\textit{$^{**}$ Spectroscopically confirmed GCs from \citet{Muller2020}, out of coverage in \citet{Haacke2025}}}\\
\footnotesize{\textit{$^{***}$ Spectroscopically confirmed GC from \citet{Haacke2025} with anomalous colours}}\\
\footnotesize{\textit{$^a$ GC candidates at 2 effective radius from the galaxy's centre or more.}}\\
\footnotesize{\textit{$^b$ GC candidates with point-like light profiles.}}\\
\footnotesize{\textit{When the \textit{u} magnitude is not provided is because the GC candidate is too faint in that band to be detected.}}
\end{table*}

\section{Ultra deep imaging of MATLAS-2019 with OSIRIS+: observational strategy and data reduction} \label{appendix:data_reduction}

Our ground-based, ultra-deep imaging of \dg{} was obtained using the instrument OSIRIS+ at the GTC. The field of view of OSIRIS+ is $7.8\arcmin \times 8.5 \arcmin$ ($7.8\arcmin \times 7.8\arcmin$ without vignetting).
The OSIRIS+ setup represents a comprehensive upgrade of the OSIRIS instrument, implemented in 2022. This upgrade includes installing OSIRIS at the Cassegrain focal station and using a new blue-sensitive monolithic detector. The camera comprises one CCD detector with a pixel scale of $0.254$ \arcsec/pixel.
In this configuration, we observed \dg{} with the $u$, $g$, and $r$ filters on four separate nights, from the 18 to 26 April 2023. In Table~\ref{tab:brightLimits}, we report the total exposure time per filter, the surface brightness limits, and the FWHM of the PSF.

\subsection{Dithering pattern}

To achieve the depths required for a detailed study of \dg{} ($\mu_{g} \sim$ 31.0 mag/arcsec$^2$, $3\sigma$ in areas equivalent to 10$\arcsec$ $\times$ 10$\arcsec$), we need an accurate estimation of the illumination of the sky at the moment of the observations. Since the twilight flats provided by the observatory are not good enough for our goals (differences between twilight illumination and illumination of the observed field), we need to use the science exposures themselves for building the flat-field and correcting the illumination of the field at once. To account for this, we followed an observational strategy similar to that used in previous works with GTC and other large telescopes (LBT, Gemini; see \citealt{TrujilloFliri2016, Trujillo_lights2021, Golini2024}). The size of the dithering step has to be similar to the size of the galaxy, so we applied a dithering scheme with a 1\arcmin{} shift and individual exposure times of $180$ seconds. In this way, the galaxy is placed at different positions in the CCD, allowing a better determination of the sky background.

\subsection{Data Reduction}
The data were downloaded from the GTC Archive\footnote{\url{https://gtc.sdc.cab.inta-csic.es/gtc/index.jsp} } and reduced using a similar procedure as in \citet{Trujillo_lights2021}. Here, we describe the main steps of the process. All the pipeline steps rely on the \texttt{Gnuastro} Software \citep{gnuastro} and its tasks, as described below.

\subsubsection{Bias and Flat field correction}

First, for each image of each filter, we masked all detector defects of the CCD. We also mask all pixels with values greater than $6.5\times 10^4~\rm{ADUs}$ to account for saturation. After this, we corrected the images from bias. To do so, we created a masterbias by combining the individual bias frames with a sigma clipping median using the \texttt{Gnuastro's task Arithmetic}. This masterbias is later subtracted from all science images.

For building the flat, we first normalize science images (bias-corrected) using the resistant mean of the values inside a ring that defines a constant-illumination section, centred on the centre of the CCD. Then, we combine the normalised, bias-subtracted science images to build a flat. However, instead of using all the images to build a master flat, we use the ones that are close in time for each image, thus allowing us to better characterize the sky illumination at the moment of the observation \citep{Saremi2025}. We define the number of frames to combine for building each of the flat-fields to be 10, characterising the field illumination on a 30 min. window. The strategy for building the flat-field is iterated a total of three times. The first flat-field is built by simply combining the images with a sigma-clipping median stacking, while on following iterations we also mask all the undetected signal before the combination. Each of this iterations improve the quality of the data, allowing us to improve the generated masks and further improve the flat itself. To avoid vignetting problems towards the corner of the detectors, we removed all the pixels where the flat-fields have a value lower than 90$\%$ the value of the central pixels.

\subsubsection{Astrometry, photometry, and sky determination}

We calculated an astrometric solution using Astrometry.net (v0.94; \citealt{Lang2010}). We used the Panoramic Survey Telescope and Rapid Response System (Pan-STARRS) DR1 \citep{panstarrs} as our astrometric reference catalogue. This produced a first astrometric solution, that we improved using SCAMP (v.2.10.0; \citealt{Bertin2006}). \texttt{SCAMP} reads catalogues generated from \sextractor{} (v.2.25.2; \citealt{Bertin1996}) and calculates the distortion coefficient of the images. After running \texttt{SCAMP}, we run \texttt{SWarp} \citep{Bertin2010} on each individual image to put them into a common grid of $3500 \times 3500$ pixels, using as resampling method LANCZOS3.

Once the astrometry is computed, we subtracted the sky of the individual frame by masking all sources using \texttt{NoiseChisel}. For each frame, we calculated the median value of the non-masked pixels and removed it from the original frame.

The calibration of the data (i.e. converting the pixel values from ADUs to physical units) is achieved by matching the fluxes of stars in our images to the fluxes of stars in an already-calibrated survey. The survey used for the calibration is \textit{The Dark Energy Camera Legacy Survey} (DECaLS) DR10 \citep{DECaLS} for the \textit{g} and \textit{r}, and the Sloan Digital Sky Survey (SDSS) DR17 \citep{SDSS2022} for the $u$-band. 

The calibration has been done as follows. First, the survey's images (DECaLs and SDSS) in the different filters of MATLAS-2019's field were downloaded, and catalogues of sources were generated. Later, in order to get a clean sample of bona fide stars, the catalogues are matched with Gaia DR3 stars. To minimise the noise introduced in the calibration we want to maximise the number of stars used, so instead of using just stars identified in Gaia, we look at the range of FWHM that these stars have, using it as a criterion for selecting point-like sources in the survey's images. Then we perform the photometry of the selected sources by using large enough apertures, which ensures that we virtually measure all the flux of the point-like source. If the aperture were too small, some flux would be lost in the tails of the PSF, generating inconsistencies when comparing fluxes measured on different images due to the differences in the PSFs. The use of a large number of stars effectively washes out the effect of the contaminants introduced by the large aperture. For defining the aperture to use, a wide range of apertures have been explored, studying at which value the measured flux is stable, concluding that apertures of $r=9R_e$ are optimal. Then the same process is applied to the data to calibrate, obtaining a catalogue of point-like sources to be used for the calibration. The apertures used for measuring the flux are $r=7R_e$. Additionally, to use non-saturated stars with high S/N, only sources with a magnitude between a certain range (16.0 < mag < 18.5, 19.5 < mag < 20.5, and 19.0 < mag < 20.0 for the \textit{u}, \textit{g} and \textit{r} bands, respectively) have been used for the calibration. Applying these criteria, we end up having between 10 to 30 stars (depending on the filter) for calibrating each of the frames.

Since the photometric calibration is a complex task, surveys have non-negligible offsets between them. Thus, in order to be able to fairly compare data calibrated with different surveys, we decide to reference our calibration to an arbitrary but common framework, this being the Gaia spectra. This introduces a correction to be performed to the fluxes measured in DECaLS and SDSS, which are compared with the fluxes measured from the Gaia spectra and corrected (by applying an offset) to match them. Secondly, in order to give a precise photometry, we need to take into account the transmittance of the filters involved in the calibration. Even if technically we are working with the same filters (in this case $u,g,r$), there are differences in the filter response between the filters of different telescopes. This can be alleviated by applying a colour correction. The colour correction applied has been characterized based on the colour ($g-r$) and is determined using the survey's filter shapes, the shapes of our filters, and GAIA spectra from stars in the field.

Finally, we match the catalogues and calculate the factor that we need to apply to our frames in order to have them calibrated to physical units. The factor computed for each frame is slightly different (due to noise and the available stars depending on the specific pointing), but since the frames to be combined in each reduction have been taken with the same configuration and parameters (same instrument, filter, exposure time, etc...) we apply the median value of the distribution of factors, reducing thus the noise in the calibration. The zero-point has been set to 22.5 in the AB magnitude system.

\subsubsection{Image co-addition and final stack}

In the final step of the data reduction, individual exposures are combined to create a final mosaic image. This step is driven by the fact that observational conditions change throughout the night, as a number of factors like the air mass, (which varies depending on the position of the target in the sky) and meteorological events like passing clouds can affect the sky brightness, degrading its quality and generating artificially high standard deviations in the sky background pixel values, indicating poorer image quality. In light of this, images captured under better conditions should carry more weight in the final mosaic to enhance the accuracy of low surface brightness features. Thus, a weighted average is used to combine the images. The weight assigned to each image is determined by the quadratic ratio of the standard deviation of the sky in the best exposure to the standard deviation of the sky in the i$^{th}$ frame (i.e. $\sigma_{min}^{2} /\sigma_{i}^{2}$, where $\sigma_{min}$ corresponds to the standard deviation of the less noisy frame). However, before stacking the data in this manner, it is crucial to mask out pixels that have been affected by undesired signals like cosmic rays. For doing this, a sigma-clipping rejection per pixel is applied using all the exposures that are going to be combined.

\begin{figure*}
    \centering
    \includegraphics[width=0.9\textwidth]{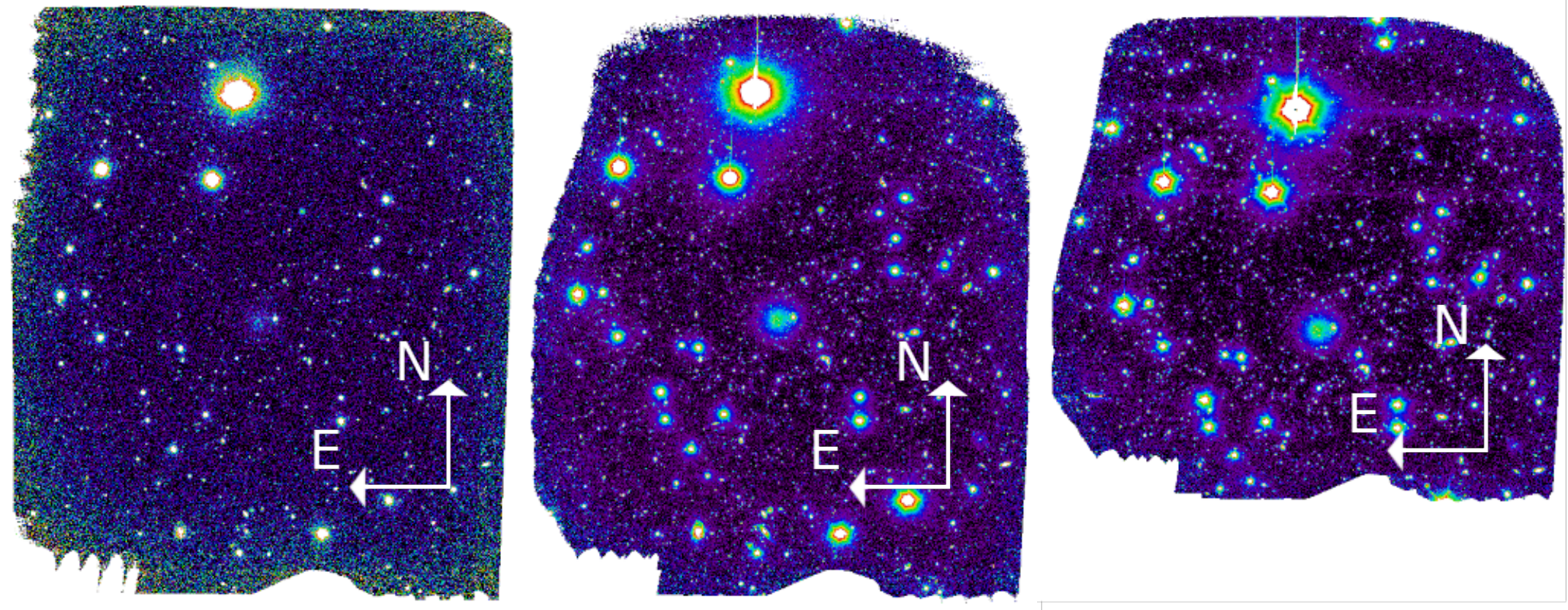}
    \caption{Entire mosaics 10$\arcmin \times$12$\arcmin$ field around \dg{} obtained with OSIRIS+. From left to right \textit{u}, \textit{g}, and \textit{r} band. \label{fig:filters}}
\end{figure*}

\begin{figure*}
    \centering
    \includegraphics[width=0.65\textwidth]{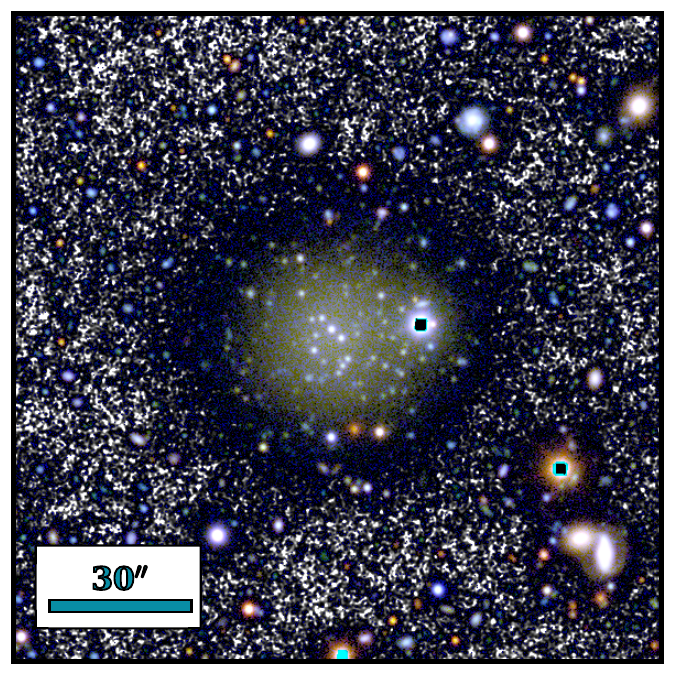}
    \caption{A region of 140\arcsec$\times$140\arcsec around \dg. The figure is a composite of an RGB image using the $u$, $g$ and $r$ OSIRIS+ bands and a black and white $g+r$ background. North is up, east is left.}\label{fig:rgbgrey}
    \label{fig:colourImage}
\end{figure*}

The co-added image is significantly deeper than any individual image and, therefore, previously invisible low surface brightness features emerge from the noise. These features subtly affect the sky determination of our individual science images, and for this reason, it is necessary to mask these regions and repeat the entire sky determination process on the individual exposures with an improved mask. In short, we repeat the sky estimation (and subsequent reduction steps) described above on the individual images using the improved masks generated by this first data co-addition. 
The mosaics for our three photometric bands are derived using the same data reduction pipeline. The final stacks have a field of view of about $10\arcmin\times 12\arcmin$, and are shown separately in Fig~\ref{fig:filters}. The colour image obtained from the final stacks is shown in \ref{fig:colourImage}.

\section{Globular cluster system asymmetry}
\label{app:asymmetry}

As pointed out in Sec. \ref{sec:SpatialDistributionOfGC}, the distribution of globular clusters of \dg{} shows an apparent asymmetry, with a larger number of candidates to the north-west. To test whether this excess of GC is statistically significant, we computed its significance using two different approaches.

First, we calculate the angular orientation of the axis that creates the maximum difference in the number of GCs between the two sides or hemispheres it separates. That results is an angle of $\sim 45^{\circ}$\footnote{With respect to the north and going from north to east.}. We divide the GCs on both sides of this axis: 23 objects in one hemisphere and 12 in the other. We then derive the probability that the GCs are distributed in this way using a binomial distribution, due to the fact that the diffuse light of the galaxy is symmetric. The probability of having a distribution of GCs as the one in \dg{} is $\sim2.4\%$. 

\begin{figure}[H]
    \centering
    \includegraphics[width=0.9\columnwidth]{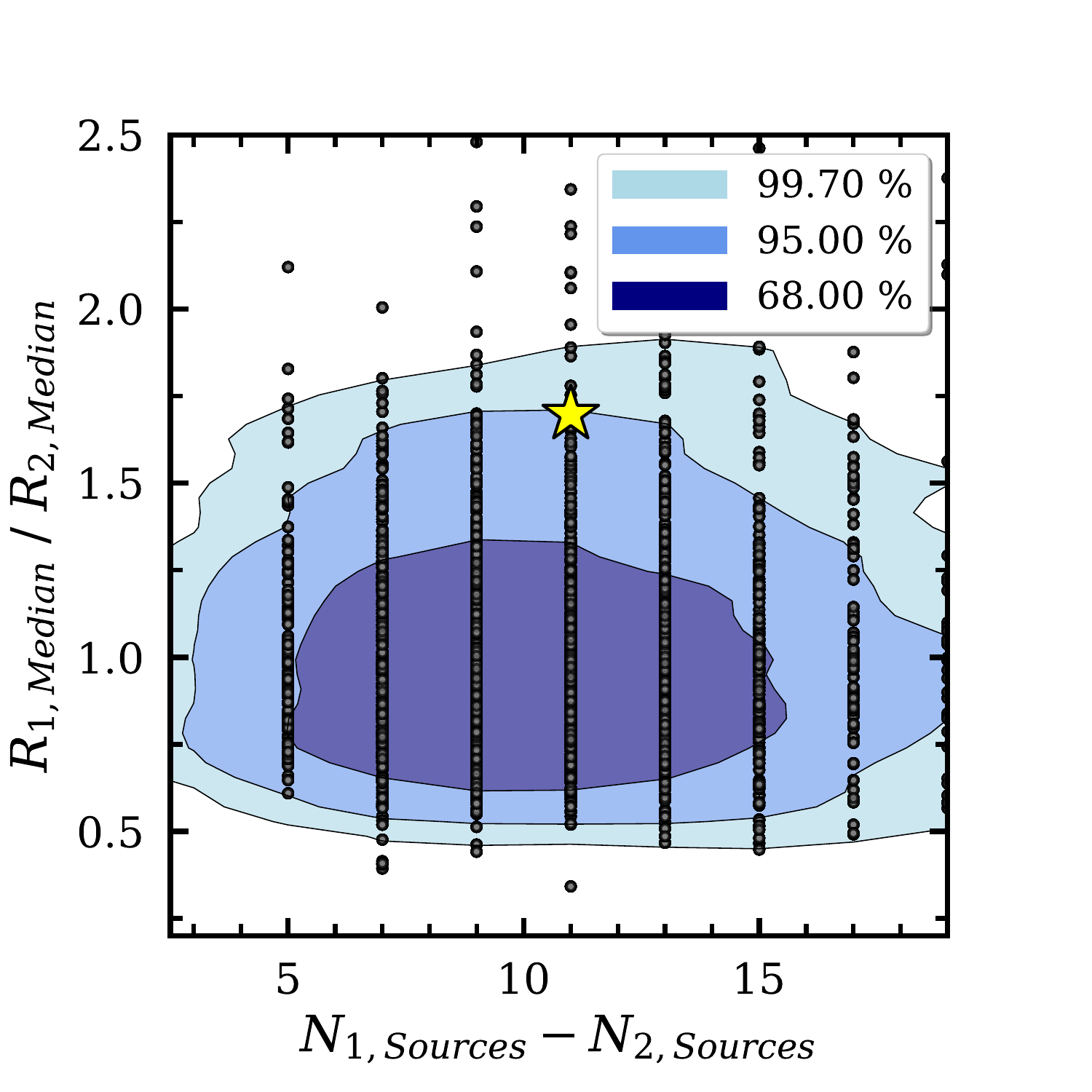}
    \caption{Probability of obtaining a GC distribution with certain parameters. The GC distribution is given by the difference in the number of GCs between two sides or hemispheres (x-axis) and the ratio between the median distance of the GCs (y-axis). The regions enclosing a 68\%, 95\% and 99.7\% of the data are indicated in different hues of blue. The star located between de 95\% and 99.7\% regions correspond to the GC distribution of \dg{}.}
    \label{fig:assymetrySim}
\end{figure}

Secondly, in addition to the difference in GCs between hemispheres, we consider the median distance from the galaxy centre to the GCs, as it also plays a role in the perceived over density of GCs. We produce $50\,000$ tests where we randomly distribute 35 GCs on the galaxy, assuming that the GC population follows the same distribution as that of the diffuse light, but reduced by a factor of $R_{e,GC}/R_e = 0.7$. In each realization, we compute the axis that maximises the difference of GCs between hemispheres (as previously done). We also compute the distance of each GC to the centre of the galaxy. Fig.~\ref{fig:assymetrySim} shows the results of the tests. We plot the difference between the number of sources in both hemispheres against the ratio of the median GC distance in each hemisphere. To compute this ratio, we assume that the side with a larger number of GCs ($N_1$) goes in the numerator ($R_1$), i.e. it is always larger than the unit. For the sake of clarity, we show the contours enclosing the 68\%, 95\% and 99.7\% of realisations. The star marker indicates the observed value for the GCs of \dg{}, falling at the boundary between the 95\% and the 99.7\% contour. In the light of this, we cannot exclude the possibility that the distribution of GCs in this galaxy is coincidental.

\section{Distance to \dg}\label{distanceToGalaxy}

\subsection{Measurement of $R_{e}$ of the GCs of \dg}
\label{app:re}
Thanks to the high angular resolution of HST, the GCs of \dg{} are partially resolved. This allows us to extract morphological information from them such as the $R_{e}$. We use \texttt{pyimfit} (a python wrapper for \texttt{imfit}, \citealt{imfit2015}) to fit of a function which approximates the intrinsic profile of a GC (a Gaussian and a King profile) convolved with the PSF of the images. We need a realistic PSF of the images, and since the PSF on space is fairly asymmetrical, the PSF derived from circularising of the 1D models obtained in Sec.~\ref{sec:psfImages} is not good enough for the measurements. Thus, to approximate the PSF in 2D, we simply use a bright, non-saturated and non-contaminated star located at RA: 15$^h$05$^m$18.66$^s$ and Dec: +1$^d$47$^m$46.61$^s$. The image used for the fits is the WFC3/\textit{F606W} image as it is the deepest. 

We perform the fits in stamps of the sources of 81 px $\times$ 81 px. The background has been estimated in the same region as for the photometry (see Sec. \ref{sec:GCSelection}), masking any neighbouring sources. 

The first intrinsic profile assumed has been a Gaussian, where the only free parameter ($\sigma$) has been allowed to vary freely. The second model explored has been the King model \citep{king1962}, which is described by a core and a tidal radius. The core radius is typically between less than one pc and a few pc, so at the proposed distances a conservative range is defined from 0.15 to 2 px. For the tidal radius, the range is between 20 and 100 pc, even though larger values are not unusual (See e.g. \citealt{king1975, Harris2002, Barmby2007}). Therefore, we only impose a lower value for the tidal radius of 2.5 px (around 10 pc). For the fits, the ellipticity was constrained between 0 and 0.2.  To sum up, the ranges for the parameters of the fits are the following:

 \begin{itemize}
    \item Gaussian prof.
    \begin{itemize}
        \item[-] $0^{\circ}$ $\leq$ PA $\leq$ $180^{\circ}$
        \item[-] 0 $\leq$ Ellipticity $\leq$ 0.2
        \item[-] $\sigma$ $\geq$ 0
    \end{itemize}

    \item King prof.
    \begin{itemize}
      \item[-] $0^{\circ}$ $\leq$ PA $\leq$ $180^{\circ}$
      \item[-] 0 $\leq$ Ellipticity $\leq$ 0.2
      \item[-] 0.15 px $\leq$ $r_c$ $\leq$ 2.0 px
      \item[-] $r_t\geq$  2.5 px
    \end{itemize}
    
\end{itemize}

We provided \texttt{pyimfit} with an error map which is the standard deviation of the background region around each source. 

\subsection{GCLF and R$_{e}$ calibrations}\label{app:dist_calibrations}

The values explored in Sec.~\ref{sec:Distance} for the peak of the GCLF are:

\begin{itemize}
  \item $M_{V, TO} = -7.66 \pm 0.09$ - MW - Metal poor GCs - \citep{Criscienzo2006}
  \item $M_{V, TO} = -7.40 \pm 0.09$ - MW - \citep{Criscienzo2006}
  \item $M_{V, TO} = -7.56 \pm 0.02$ - dIrr - \citep{Georgiev2009_2}
\end{itemize}

The conversion for MATLAS-2019's globular clusters magnitudes between Vega and AB is $V_{606}(AB)$ = $V(Vega)$ - 0.145 mag. This is derived computing $V_{606}(AB)$ and $V$(Vega) of a simple stellar population model from \citet{Vazdekis2016} with [Fe/H] = $-1.44$ and age 10 Gyr.\\

The values explored for the peak of the $R_{e}$ distribution are:

\begin{itemize}

  \item $Log(R_{e, TO}) = 0.520 \pm 0.0236$ pc - GCs from Dwarfs 
  \item $Log(R_{e, TO}) = 0.506 \pm 0.0239$ pc - GCs from MW

\end{itemize}

The estimates for the $R_{e}$ have been obtained directly from the catalogues \citep{Harris1996, Georgiev2009}, using a 3$\sigma$-clipped mean after transforming the radii to logarithmic scale. The errors given are the errors of the mean value of the distribution, (i.e. the standard errors).

\subsection{Comparison with local GCs and MATLAS survey dwarfs GCs}\label{app:comparisongcs}

\begin{figure}
    \centering
    \includegraphics[width=0.9\linewidth]{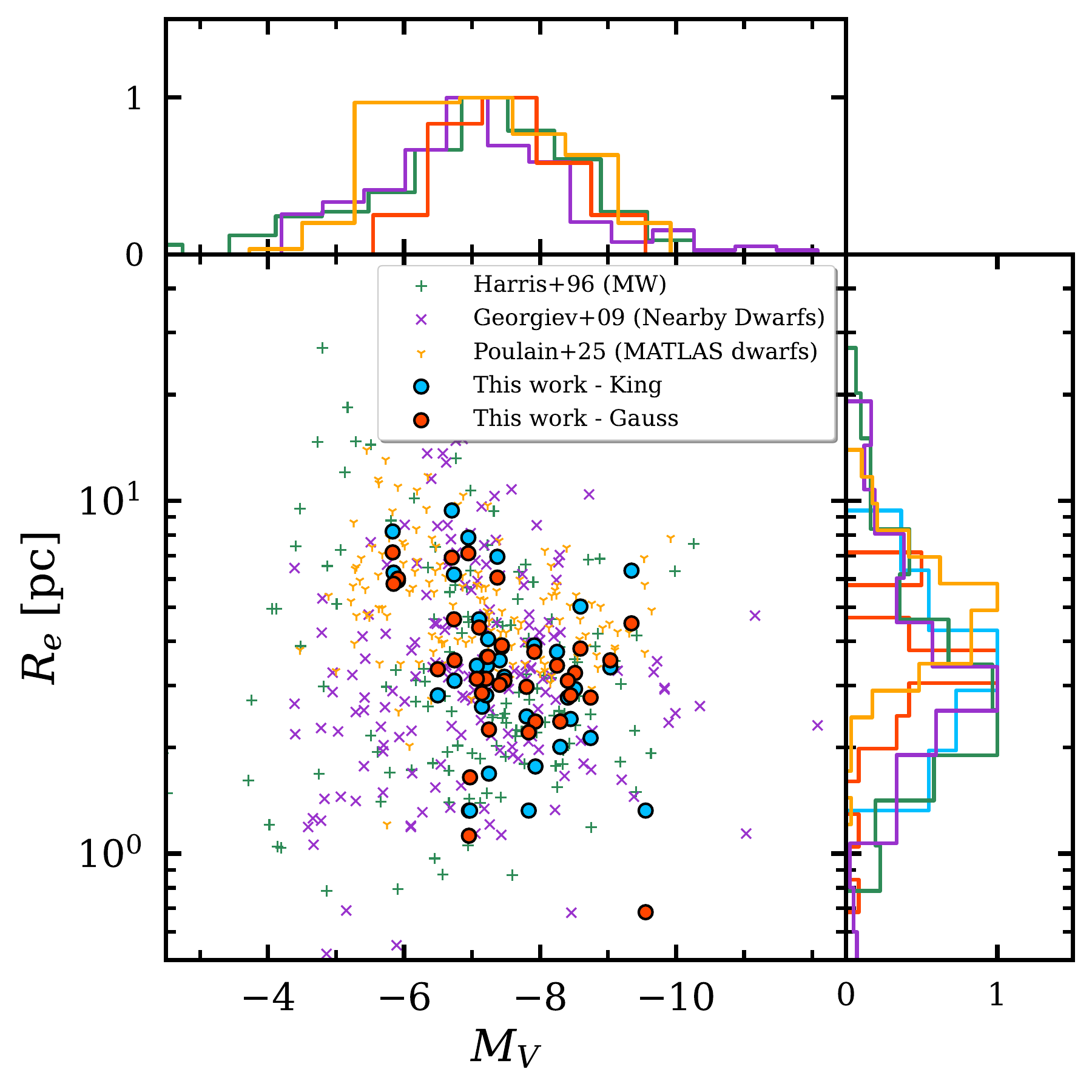}
    \caption{Comparison between the $M_V$ and $R_e$ of the GC population of \dg{} with the GCs from the Milky Way \citep{Harris1996}, from nearby dwarf galaxies \citep{Georgiev2009} and from dwarf galaxies in the MATLAS survey \citep{poulain2025}.  The assumed distance to \dg{} is 20.7 Mpc.}
    \label{Fig:DistanceComparisonFirst}
\end{figure}

\begin{figure}
    \centering
    \includegraphics[width=0.9\linewidth]{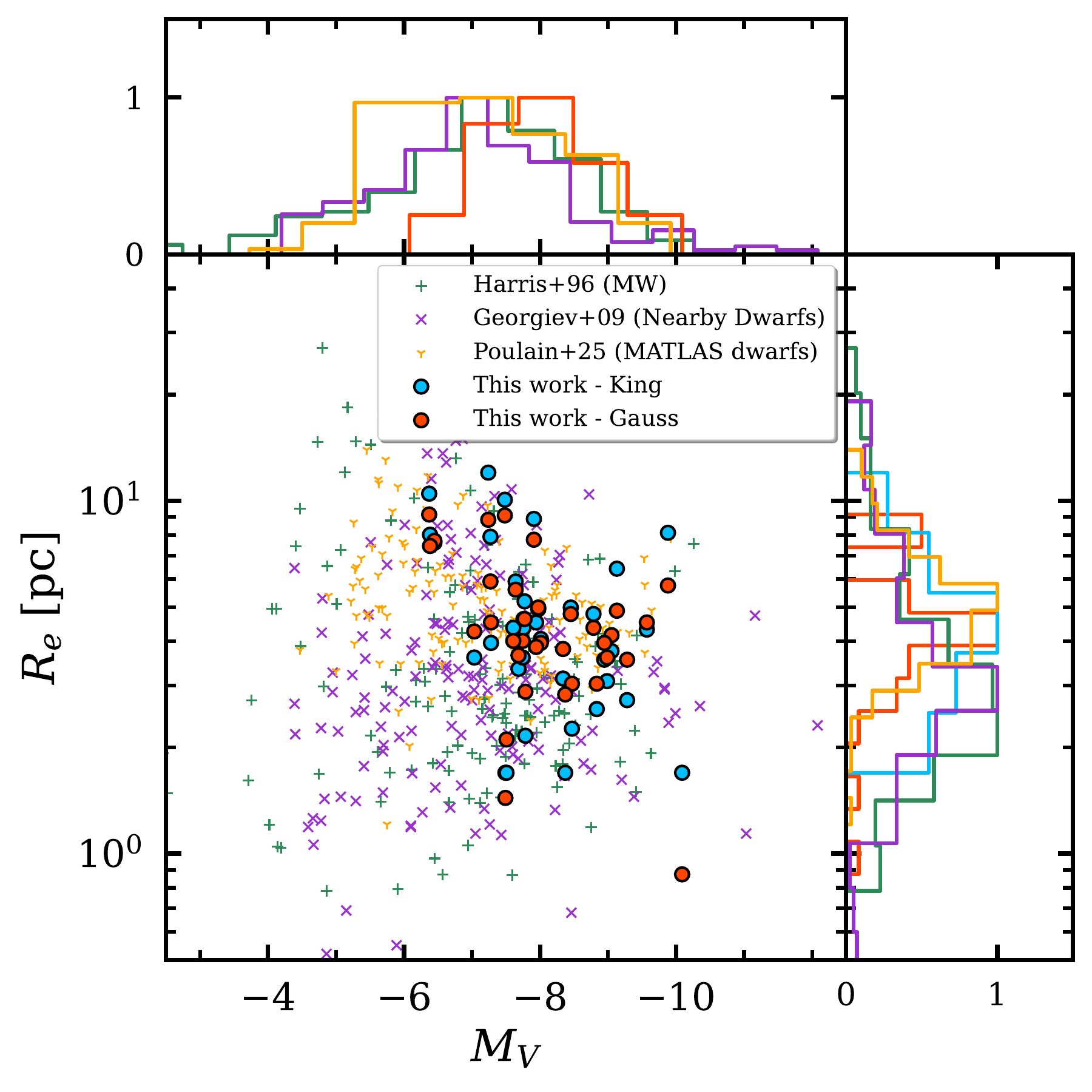}
    \caption{Same as Fig. \ref{Fig:DistanceComparisonFirst} but assuming a distance to \dg{} of 26.5 Mpc.}
    \label{Fig:DistanceComparisonSecond}
\end{figure}

We compared the distribution of magnitudes and $R_{e}$ of the GCs of \dg{}, at the two possible distances of the galaxy, with the properties of local MW GCs \citep{Harris1996}, GCs from nearby dwarfs \citep{Georgiev2009}, and of GCs from dwarf galaxies from the MATLAS survey \cite{poulain2025}. These are shown in Figures \ref{Fig:DistanceComparisonFirst} at 20.7 Mpc and \ref{Fig:DistanceComparisonSecond} at 26.5 Mpc.\\

Looking at the local sample we see how magnitudes and effective radius seems not to show any trend, and how most of them occupy the region between $-4.0\, < M_v < -11\,mag$ and $0.5\, < R_e < 10\,pc$. The comparison of MATLAS-2019 GC population with the locals allows us to visually inspect how alike are the populations depending on the distance.

\end{appendix}
\end{document}